\RequirePackage{lineno}
\documentclass[aps,prl,twocolumn,showpacs,superscriptaddress,groupedaddress]{revtex4}  % for review and submission
\usepackage{graphicx}  % needed for figures
\usepackage{dcolumn}   % needed for some tables
\usepackage{bm}        % for math
\usepackage{amssymb}   % for math
\usepackage{color}
\usepackage{ulem}
\usepackage[bookmarks=true]{hyperref}

\newcommand{\cor}[1]{\textcolor{blue}{\textbf{#1}}}

\begin{document}

\setpagewiselinenumbers
\modulolinenumbers[1]
%\linenumbers

%\title{Multi-Reactor and Multi-Detector Experiments Reactor Flux Systematics}
\title{Reactor Neutrino Flux Uncertainty Suppression on Multiple Detector Experiments}

%\email{{\tt cucoanes@subatech.in2p3.fr} and {\tt anatael@in2p3.fr}}

\author{A.~S.~Cucoanes$^{2}$\footnote{Corresponding:~{\tt cucoanes@subatech.in2p3.fr}}}
\noaffiliation{} %SUBATECH, CNRS/IN2P3, Universit\'e de Nantes, Ecole des Mines de Nantes, F-44307 Nantes, France}

\author{P.~Novella}
\affiliation{APC, Astro-Particule et Cosmologie, CNRS/IN2P3, Universit\'e Paris Diderot, 75205 Paris Cedex 13, France}
%\email[Corresponding authors:]{ {\tt cucoanes@subatech.in2p3.fr} and {\tt anatael@in2p3.fr}.}

\author{A.~Cabrera$^{1}$\footnote{Corresponding:~{\tt anatael@in2p3.fr}}}
\noaffiliation{} %APC, Astro-Particule et Cosmologie, CNRS/IN2P3, Universit\'e Paris Diderot, 75205 Paris Cedex 13, France}
%\email[Corresponding authors:]{ {\tt anatael@in2p3.fr}}

\author{M.~Fallot}
\affiliation{SUBATECH, CNRS/IN2P3, Universit\'e de Nantes, Ecole des Mines de Nantes, F-44307 Nantes, France}

%\author{H.~de~Kerret}
%\affiliation{APC, Astro-Particule et Cosmologie, CNRS/IN2P3, Universit\'e Paris Diderot, 75205 Paris Cedex 13, France}

\author{A.~Onillon}
\affiliation{SUBATECH, CNRS/IN2P3, Universit\'e de Nantes, Ecole des Mines de Nantes, F-44307 Nantes, France}

\author{M.~Obolensky}
\affiliation{APC, Astro-Particule et Cosmologie, CNRS/IN2P3, Universit\'e Paris Diderot, 75205 Paris Cedex 13, France}

\author{F.~Yermia}
\affiliation{SUBATECH, CNRS/IN2P3, Universit\'e de Nantes, Ecole des Mines de Nantes, F-44307 Nantes, France}
\date{\today}

\begin{abstract}
\noindent
This publication provides a coherent treatment for the reactor neutrino flux uncertainties suppression, specially focussed on the latest $\theta_{13}$ measurement. The treatment starts with single detector in single reactor site, most relevant for all reactor experiments beyond $\theta_{13}$. We demonstrate there is no trivial error cancellation, thus the flux systematic error can remain dominant even after the adoption of multi-detector configurations. However, three mechanisms for flux error suppression have been identified and calculated in the context of Double Chooz, Daya Bay and RENO sites. Our analysis computes the error {\it suppression fraction} using simplified scenarios to maximise relative comparison among experiments. We have validated the only mechanism exploited so far by experiments to improve the precision of the published $\theta_{13}$. The other two newly identified mechanisms could lead to total error flux cancellation under specific conditions and are expected to have major implications on the global $\theta_{13}$ knowledge today. First, Double Chooz, in its final  configuration, is the only experiment benefiting from a negligible reactor flux error due to a $\sim$90\% geometrical suppression. Second, Daya Bay and RENO could benefit from their partial geometrical cancellation, yielding a potential $\sim$50\% error suppression, thus significantly improving the global $\theta_{13}$ precision today. And third, we illustrate the rationale behind further error suppression upon the exploitation of the inter-reactor error correlations, so far neglected. So, our publication is a key step forward in the context of high precision neutrino reactor experiments providing insight on the suppression of their intrinsic flux error uncertainty, thus affecting past and current experimental results, as well as the design of future experiments.

\end{abstract}

\maketitle

\section{Introduction}

Reactor neutrinos have been used for fundamental research since the discovery of neutrinos. 
The last decade have witnessed a remarkable reduction of systematic error in reactor neutrino experiments, by about one of order of magnitude, imposed by the high precision needed to measure $\theta_{13}$ by Double Chooz~\cite{Abe:2013sxa}~(DC), Daya Bay~\cite{DB2,DB3}~(DB) and RENO~\cite{Ahn:2012nd}, a milestone for the world strategy of neutrino flavour research.
RENO has released several updates of the $\theta_{13}$ analysis in conferences here disregarded until publications follow.
The reactor measurements are consistent with all measurements sensitive to $\theta_{13}$~\cite{T2Kextra, MIextra} obtained via other techniques providing a coherent $\theta_{13}$ perspective, as obtained by global fit analyses~\cite{GFA1extra, GFA2extra, GFA3extra}. 
Since the reactor $\theta_{13}$ experiments precision is unrivalled, they are expected to dominate the world knowledge on $\theta_{13}$, likely, for a few decades to go.
Hence, reactor systematic dominates much of the $\theta_{13}$ world knowledge, as experiments reach their final sensitivities.
% since no improvement is foreseen.
The measured $\theta_{13}$ (and its uncertainty) is expected to play a critical role to constrain, or measure, still unknown neutrino oscillation observables, such as CP-violation and the atmospheric mass hierarchy~\cite{prospect}.

To maximise the sensitivity to $\theta_{13}$, reactor experiments were forced to conceived experimental setups where flux, detection and background systematics are controlled to the unprecedented level of a few per-mille each contribution (\textless1\% total).
The statical resolution is boosted by using multi-reactors sites.
The unprecedented precision achieved is experimentally very challenging, therefore the redundancy $\theta_{13}$-experiments is critical, specially if their uncertainty budgets are complementary to offer maximal cross-validation.
Despite some complementary, the reactor experimental setups are unavoidably similar and suffer from similar limitations, hence validation by different techniques would be important, although the precision needed is unattainable today.
The fore-mentioned precision improvement was obtained via multi-detector experimental setups, whereby, at least, two detectors are used for the reduction of the overall systematic budget since correlated systematics among detectors cancel out.
This way, while the {\it absolute systematics} are the same, the {\it relative systematics} are much lower.
%, allowing today's precision on $\theta_{13}$.
The {\it absolute systematics} are still dominant in any single-detector setup, such as DC (single detector), but also all past and, likely, most future reactor experiments.
%, where DC has achieved so far most precise results.

\begin{figure*}[htb!]
\centering
\includegraphics[scale=0.65]{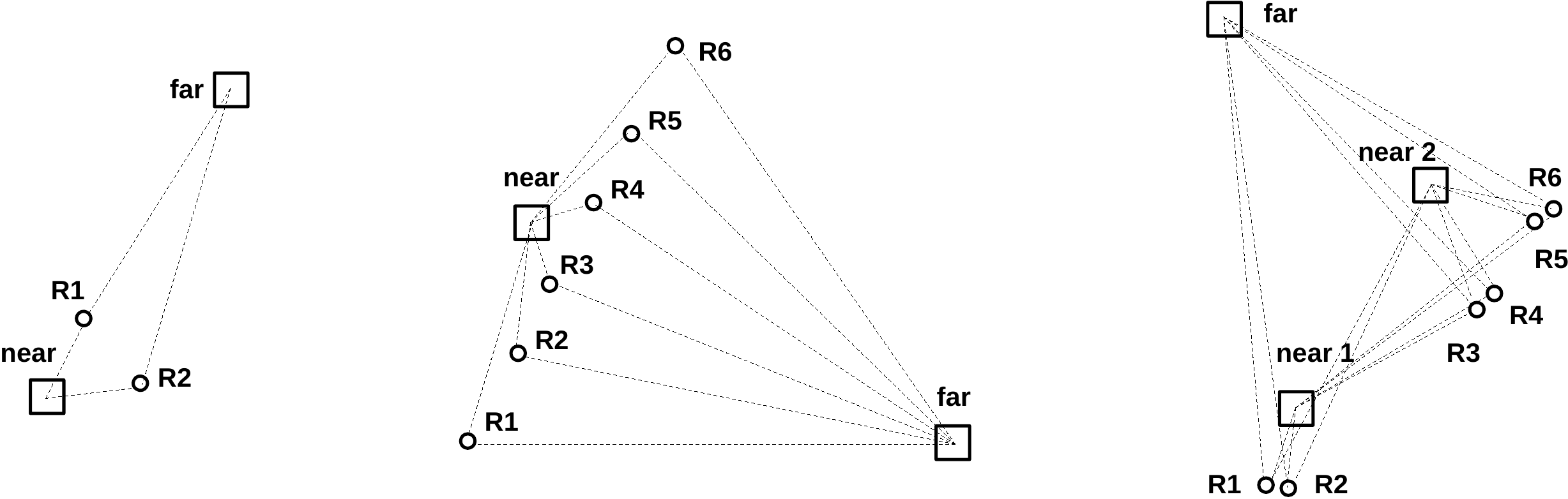}
\caption{ 
	The 2D geometry of the experimental setups of Double Chooz (left), RENO (middle) and Daya Bay (right) experiments is shown. 
	The squares indicate the detectors and the circles indicate the reactor cores. 
	The dotted lines depict the baselines between detectors and reactors, while distances are summarised in Table~\ref{Tab-distances}. 
	Notation on Daya Bay: AD1, AD3 and AD4 correspond to the here called near-1, near-2 and far, respectively. 
	}
\label{Fig-Expsetups}
\end{figure*}

The systematics reduced by the multi-detector configuration are: detection and flux systematics.
Detection systematics benefit from dedicated detector design for them to be {\it effectively identical}, typically so, only upon full calibration, thus implying the same (or negligibly different) responses and composition (cross-section, proton number, etc). 
Flux systematics benefit from the fact that the {\it near} detector(s) is located closer the reactor(s) such that the flux modulation originating from neutrino oscillations, $\theta_{13}$ in this case, is negligible (or very small) compared to the {\it far} detector(s) located further away.
The far is placed at the expected maximal oscillation deficit driven by the $\Delta m^2$ (atmospheric) constraint by MINOS~\cite{MIextra} and T2K~\cite{T2Kextra}.
The suppression of the flux systematic is highly non-trivial, being the main subject of this publication.
While having effectively identical detectors suffices to reduce detection systematics, from $\sim$2.0\%~\cite{DB2} to $\sim$0.2\%~\cite{DB2}, just having near and far detectors will not necessarily provide a full cancellation of flux systematics - unlike what was originally thought.
There are three mechanisms leading to possible flux systematic reduction to be elaborated in detail within the paper.
First, multi-reactor uncorrelated uncertainties can benefit from having several identical reactors.
Second, the near-far geometry of the experimental setup could enhance ability for the near become an effective {\it perfect monitor} to the far; i.e. cancelling fully the flux systematic error.
And, third, the nature of the reactor uncertainties; i.e. whether correlated or uncorrelated among reactors; might be exploited, as measured by multiple detectors.
% since those errors trivially propagate with $1/\sqrt{R}$, where R stands for the number of reactors in the experimental site. 
Reactor systematics suppress from single detector scenario, typically $\sim$3\%~\cite{DB2} to \textless1\% for multi-detector setups.
DC achieves an impressive $\sim$1.7\%~\cite{Abe:2013sxa} for a single-detector setup using the Bugey4 data for the mean cross-section per fission normalisation of the fission; i.e. as an effective {\it normalisation-only near}.
The most accurate reactor flux anti-neutrino spectrum predictions~\cite{spectra1,spectra2} rely on the ILL uranium and plutonium isotopes input data~\cite{ILL1,ILL2}.
However, the final flux systematics are a combination of the reactor and spectral systematic errors and depends on depends on each experiment configuration, since the evolution of the fission elements depends on the running configuration of each reactor.
Therefore flux systematics are expected, with current knowledge, to be the dominant contribution for some of the reactor $\theta_{13}$ experiments, such as DB, itself leading the world $\theta_{13}$ precision.
% - a remarkable achievement on the control of all other systematics.

This publication develops a framework for the non-trivial propagation of the reactor flux uncertainties and their suppression in the context of multi-detector and multi-reactor experimental setups, providing mechanism for the improvement of the global $\theta_{13}$ precision.
%, most relevant for reactor $\theta_{13}$ experiments, as they reach their final sensitivities.
%The discussion here presented, in addition, sheds light into some relevant issues to be considered for the combination of reactor-$\theta_{13}$ measurements, as envisaged by the three collaborations. 
Most our discussion stays generic on flux systematic propagation; i.e. no need for the specific experiment error breakdown, allowing easy relative comparison across all experiments.
Our study cases, however, inspired on the specific reactor-$\theta_{13}$ experiments configurations for maximal pertinence.
The core of our calculations is analytical, but a cross-check was implemented using a dedicated Monte-Carlo-based analysis, is also presented.
The discussion starts from the simplest single detector configuration,
%, most relevant to most experiments in the world, beyond the specific case of $\theta_{13}$ experiments.
then evolving towards a general formalism applicable to any multi-detector and multi-reactor setups.
The numbers linked to the specific experimental setups are however only guidelines as they are obtained in simplified scenarios, again, to maximise comparability.

%%%%%%%%%%%%%%%%%%%%%%%%
\section{Flux Induced Uncertainties}
%%%%%%%%%%%%%%%%%%%%%%%%

So far, the reactor $\theta_{13}$-experiments appear to treat the reactor systematics similarly.
However, all collaborations are lacking dedicated publications on the topic, so questions about the coherence across collaboration remain.
Below, we will briefly summarise the reactor flux systematics information, as provided by the different collaborations.
Most experiments rely on PWR reactors, hence the anti-neutrino flux is dominated by four isotopes: $^{235}$U, $^{239}$Pu,$^{241}$Pu and $^{238}$U (ordered by contribution relevance).
%the most conservative way; i.e. maximising their flux uncertainties.

%REACTOR ERRORS
In general, flux systematics uncertainties are typically broken down into three terms: thermal power, fission fractions and spent fuel, as summarised in Table~\ref{Tab-uncorpar}.
The thermal power history is usually provided by the electricity company exploiting the reactor.
The associated error is related to the measurement method, the precision of the installed sensors and the employed sensors calibration~\cite{Djurcic}.
Each experiment has to precisely study the way how the thermal power is measured in order to estimate the corresponding error and the possible correlations among all the reactors involved, as typically, the same measurement techniques are applied to all identical reactors in a power plant.
The fractional fission rates have to be computed through reactor simulations, while their uncertainty estimation is not simple as they depend on the reactor model and the approximations used for each setup~\cite{Onillon}.
After each cycle, the reactors are stopped for refuelling, typically once per year in a few weeks lasting operation. 
In this operation, part of the fuel assemblies are exchanged with new ones. 
The spent fuel are stored in the dedicated pools, typically located next to the reactor sites with slightly different baselines.
The spent fuel long lived fission products can generate a small fraction of anti-neutrinos above the inverse beta decay threshold~\cite{DBspent}. 
Therefore, each experiment has to evaluate also the contribution from spent fuel and its uncertainty.

\begin{table}
\caption{
	Reactor-$\theta_{13}$ experiments breakdown of detector flux 1$\sigma$ systematic error~\cite{Abe:2013sxa, DB2, Ahn:2012nd}.
	The Daya Bay and RENO errors quoted here are correspond to their uncorrelated contributions, as they relied on multi-detector setups.
	Instead, the Double Chooz errors includes also the correlated contributions, as quoted from single-detector analysis. 
	Therefore, the Double Chooz error is expected to be overestimated and will be revised by the collaboration in future publications.   
	Thermal power ($P_{th}$), fission fractions ($\alpha_f$) and spent fuel are considered.	
	}
\begin{ruledtabular}
\begin{tabular}{lcccc}
             & $P_{th}$ & $\alpha_f$ & Spent Fuel & Total \\
             & (\%)     & (\%)       & (\%)       & (\%)  \\
\hline
Double Chooz & 0.5      & 0.9        & included   & 1.0   \\
Daya Bay     & 0.5      & 0.6        & 0.3        & 0.8   \\
RENO         & 0.5      & 0.7        & unknown    & 0.9   \\
\end{tabular}
\end{ruledtabular}
\label{Tab-uncorpar}
\end{table}

DC, until now operating in single-detector mode (DC-I), quotes its reactor uncertainties as fully correlated between reactors, which is the most conservative approach for this configuration as it will be demonstrated later on.
%illustrated in Fig.~\ref{Fig-DC_0near_2D}.
The values for the thermal power and the fission fraction are 0.5\% and 0.9\%, respectively~\cite{Abe:2013sxa}.
The spent fuel contribution was estimated within the dominant 1.7\% flux systematic error quoted for all DC-I publications.
The DC-I errors quoted account for both correlated and uncorrelated contributions, thus they are overestimated relative to its multi-detector configuration where only the uncorrelated errors are to be considered, as quoted by Daya Bay and RENO.
DC is expected revise these numbers in forthcoming the multi-detector DC (data taking started in 2014).
%Therefore, in this publication, this will make the DC to be look larger than expected, but as it will be seen, 
However, our description handles error suppression in relative, thus avoiding absolute error estimation and/or premature discussion, left for the publications by the experiments.
%Thus, our results can be adapted easily to future expected revisions.
%However, a explicit statement is expected for forthcoming DC-II results.

DB and RENO, instead, quote systematics for their multi-detector configurations; i.e. only the detector uncorrelated components.
%, assuming they are fully uncorrelated among their reactors; i.e. no reduction possible, as illustrated in Fig.~\ref{Fig-DC_1near_2D} for the case of DC-II.
DB quotes, as thermal power and fission fraction systematics, 0.5\% and 0.6\% respectively~\cite{DB2}.
An extra contribution from the spent fuel is estimated to be 0.3\%, increasing the overall systematic to 0.8\%~\cite{DB2}.
DB presented a reduction of the total systematic uncertainty to 0.04\% on its ratio between observed over predicted rate upon optimisation, allowing modulation of the different near contributions relative to the far.
This analysis, although not used for the measurement of $\theta_{13}$, raises unsettled debate about the physical meaning of such a modulation, a priori fixed by the geometrical configuration of the site.
For completeness, we have replicated such a result in this publication, however we disregard it fully for discussion on the precision of $\theta_{13}$.
RENO, on the other hand, quotes 0.5\% and 0.7\%~\cite{Ahn:2012nd}, respectively, for the thermal power and fission fractions, leading to an overall total reactor uncorrelated flux systematic uncertainty of 0.9\%, in consistent agreement with DB. 
RENO quotes no error for the spent fuel, so it is neglected hereafter, however this is somewhat unexpected since its contribution is expected to be non-negligible given the large number of reactors, like in the case of DB.

\begin{table*}
\caption{
	The distances in meters between detectors and reactors used for calculation, as illustrated in Fig~\ref{Fig-Expsetups}. 
	In the cases of DC and RENO, we inferred from~[\onlinecite{Abe:2013sxa,Ahn:2012nd}], while for DB the information is clearly provided in~\cite{DB2}.
	}
\begin{ruledtabular}
\begin{tabular}{lcccccc}
Setup                    & $R_1$  & $R_2$  & $R_3$  & $R_4$  & $R_5$  & $R_6$   \\ 
\hline
{\bf Double Chooz}\\ 
\hspace{0.5cm} far   & 997.9  & 1114.6 &        &        &        &         \\
\hspace{0.5cm} near  & 351    & 466    &        &        &        &         \\
\hline
{\bf Daya Bay}\\
\hspace{0.5cm} far       & 1920   & 1894   & 1533   & 1534   & 1551   & 1525    \\
\hspace{0.5cm} near-1    & 362    & 372    & 903    & 817    & 1354   & 1265    \\
\hspace{0.5cm} near-2    & 1332   & 1358   & 468    & 490    & 558    & 499     \\
\hline
{\bf RENO} \\
\hspace{0.5cm} far           & 1556.5 & 1456.2 & 1395.9 & 1381.3 & 1413.8 & 1490.1  \\
\hspace{0.5cm} near          & 667.9  & 451.8  & 304.8  & 336.1  & 513.9  & 739.1   \\
\end{tabular}
\end{ruledtabular}
\label{Tab-distances}
\end{table*}

%%%%%%%%%%%%%%%%%%%%%%%%
\section{The Iso-Flux Condition}
%%%%%%%%%%%%%%%%%%%%%%%%

An important consideration for the reduction of reactor flux systematics is the geometry of the experimental setup, illustrated in Fig.~\ref{Fig-Expsetups}, which might lead to total error cancellation.
The location of the detector strongly depends on the overburden topology of the site, critical for background suppression.
In a multi-detector experimental site configuration, the flux uncertainties would cancel entirely, regardless of the nature of the uncertainties, if the relative contribution made by each given reactor to the total detected antineutrino flux is the same for all the detectors.
This is a simple acceptance condition, as it will be demonstrated later on.
If met, the near becomes an effective {\it perfect monitor} of the far, thus the flux systematic uncertainty is null.
This condition is, in fact, trivially possible in case of isotropic sources, like reactors and, unfortunately, typically impractical in meson-decay neutrino induced beams (i.e. decay in flight), where laborious dedicated efforts, including dedicated experiments, are needed for an accurate spectral extrapolation from near to far, like in the case of MINOS and T2K.
Complex multi-reactor sites experiments might not meet iso-flux condition.
%, which is the case of DB and RENO.
However, the iso-flux condition might be fulfilled partially, thus yielding partial suppression of systematics.
%that is to be evaluated carefully by each experiment, as it can be exploited to reduce systematics.
%Although, DC is expected to be almost iso-flux, t
As the iso-flux condition is not fully fulfilled by any of the reactor-$\theta_{13}$ experiments, the estimation of the reactor systematics is not straightforward.
So, it is mandatory to account for the differences among the reactor compositions fluxes and their corresponding systematic errors. 
%Considering an individual reactor, the main contributions to the total uncertainty are given by the systematic errors associated to the thermal power, the core composition and the spent fuel. 
%These information may be determined through different methods, depending on each experiment.
In fact, this will be investigated and quantitatively estimated in the following sections for each experiment, using some simple approximations.

%%%%%%%%%%%%%%%%%%%%%%%% 
\section{Reactor Uncertainty Suppression}
%%%%%%%%%%%%%%%%%%%%%%%%

Let us consider a general experimental setup where one or more detectors D measure antineutrino fluxes generated by $N_R$ reactors. 
For each detector, the incoming flux $\Phi_D$ represents a superposition of the fluxes emitted by $N_R$ reactors having the magnitudes $\Phi_R^i$ weighted by the corresponding solid angles $\Omega_D^i$ subtended between the detector and each reactor. 
We express this as

\begin{equation}
\Phi_D = \sum_{Ri=R1}^{N_R} \Omega_{DRi} \Phi_{Ri}
\label{eq:defFIgen}
\end{equation}

\noindent where index $Ri$ runs over all the reactors (R1, R2,\dots,$N_R$) per site and the index $D$ stands for the detector(s), where the near(s) and far(s) will be indicated later on by $n$ and $f$, respectively.

The accuracy and the precision of the geometrical solid angles is carefully controlled by the experiments by dedicated surveys. 
The relevant distances are summarised in Table~\ref{Tab-distances}. 
The associated uncertainty is generally so small relative to other terms, that we can neglect them for our analysis. 
We can also neglect the effect of neutrino oscillation, as we consider the un-oscillated integral flux prediction.

Before going through the specific cases, let us define three quantities to be used throughout the formalism.
They characterise the systematic error associated to the emitted anti-neutrino flux contributing to the error of the measurement made by detectors.
$\delta_{Ri}$ represents the relative error of the emitted anti-neutrino flux per single reactor core $Ri$; i.e. the relative error of $\Phi_{Ri}$.
$\delta_D$ is the flux reactor uncertainty, as measured by the experiment; i.e. the relative error of $\Phi_{D}$.
For an experimentalist considering a precise measurement of $\theta_{13}$, the most important quantity is $\delta_D$, but the raw input is $\delta_R$. 
The ratio between $\delta_D$ over $\delta_R$ provides a measure for the effective error {\it suppression fraction} (SF) obtained upon error propagation, which is the core parameter used for error suppression characterisation.
This factor reflects the ability of each experiment to reduce the overall reactor uncertainty relative to the simplest case of {\it one detector with one reactor} configuration (i.e. $\delta_R$), where no cancellation is expected.
The convention use is that the smaller the SF, the smaller the final reactor flux error systematic as measured at the detector ($\delta_D$), so SF is ranges within the interval [0,1].
The extreme values 0 and 1 stand for total suppression ($\delta_D = 0$, regardless of $\delta_R$) and no suppression ($\delta_D = \delta_R$), respectively.

In this section, our formalism focuses on the propagation of the flux uncertainties, as characterised by the effective SF starting from Eq.~\ref{eq:defFIgen} as applied to the specific configuration of each setup and considering the same contribution for each individual reactor and ignoring the nature of each error component.
Hence, most of our description is done in relative; i.e. no necessity to introduce absolute error discussion; providing the ideal framework for highlighting the main features of each experiment in a comparable basis. 
As convention, we took the actual experimental setups (next sub-sections) to illustrate our discussion.
However, most of the discussion remains generic and figures will aim to be exemplify the general features.
To simplify the description flow, some calculations are only shown in the Appendix.

%Finally, the critical evaluation of the different components of the total flux uncertainty budget is addressed in the discussion section at the end, as it does not affect the generic development shown in this section.
%In addition, very delicate considerations are to be contemplated for complete discussion of this subject, some will be addressed and some exceed even the scope of this publication, as they can only be addressed by the specific experiments upon quoting their final errors on their $\theta_{13}$ measurement.
%on the input errors published and used by each experiment, prior to aiming to obtain, upon the here reported potential suppression, a 

%%%%%%%%%%%%%%%%%%%%%%%%
\subsection{Single Detector: the DC-I Configuration}
%%%%%%%%%%%%%%%%%%%%%%%%

This case is best illustrated by the DC-I (far only phase), where $N_R$ stands for 2 reactors, R1 and R2.
Propagating the errors on $\Phi_D$ (Eq.~\ref{eq:defFIgen}) for the sole detector involved, we obtain the corresponding SF expressed as

\begin{equation} 
\mathrm{SF}^2 = 1 - \frac{ 2 \Omega_{R1} \Phi_{R1} \Omega_{R2} \Phi_{R2} (1-k) }{ (\Omega_{R1} \Phi_{R1} + \Omega_{R2} \Phi_{R2})^2 } 
\label{eq:defI1D2R}
\end{equation} 

\noindent 
where k stands for the {\it correlation factor} between the generated fluxes defined (see Eq.~\ref{eq_kdeb}) as

\begin{equation}        
k = \frac{1 + \frac{\delta_R^c - \delta_R^u}{\delta_R^c + \delta_R^u} }{\sqrt{2 \left[1 + \left( \frac{\delta_R^c - \delta_R^u}{\delta_R^c + \delta_R^u} \right)^2 \right]}}
\label{eq:kdef}
\end{equation}

\noindent 
where $\delta_R^c$ and $\delta_R^u$ are, respectively, the {\it correlated} and the {\it uncorrelated} components of the total reactor flux uncertainty per individual core, defined as $(\delta_R)^2 = (\delta_R^c)^2 + (\delta_R^u)^2$. Then, we can rewrite Eq.~\ref{eq:defI1D2R} as

\begin{equation} 
\mathrm{SF}^2 = \frac{ 1 + \Omega_{R}^2 + 2 k \Omega_{R}}{ (1 + \Omega_{R})^2 } 
\label{eq_finI1D2R}
\end{equation} 

\noindent 
using a term proportional to the difference between the reactor fluxes weighted by the corresponding solid angles

\begin{equation} 
\Omega_{R} = \frac{\Omega_{R2} \Phi_{R2}}{\Omega_{R1} \Phi_{R1}} \cong \frac{L_{R1}^2 \Phi_{R2}}{L_{R2}^2 \Phi_{R1}}     
\label{eq:defOMR}
\end{equation} 

\noindent 
where $L_{R1}$ and $L_{R2}$ are the distances between the detector and the reactors. The proportionality between the solid angles and the corresponding detector-reactor distances is given by the inverse-square law. The results provided by Eq.~\ref{eq_finI1D2R} remain the same for a different choice of the reference reactor in Eq.~\ref{eq:defOMR}, more precisely considering the ratio $(\Omega_{R1} \Phi_{R1})/(\Omega_{R2} \Phi_{R2})$. Indeed, this case is equivalent to a transformation $\Omega_{dR} \to 1/\Omega_{dR}$ which leaves Eq.~\ref{eq_finI1D2R} unchanged. 

\begin{figure}[htb!]
\centering
\includegraphics[scale=0.45]{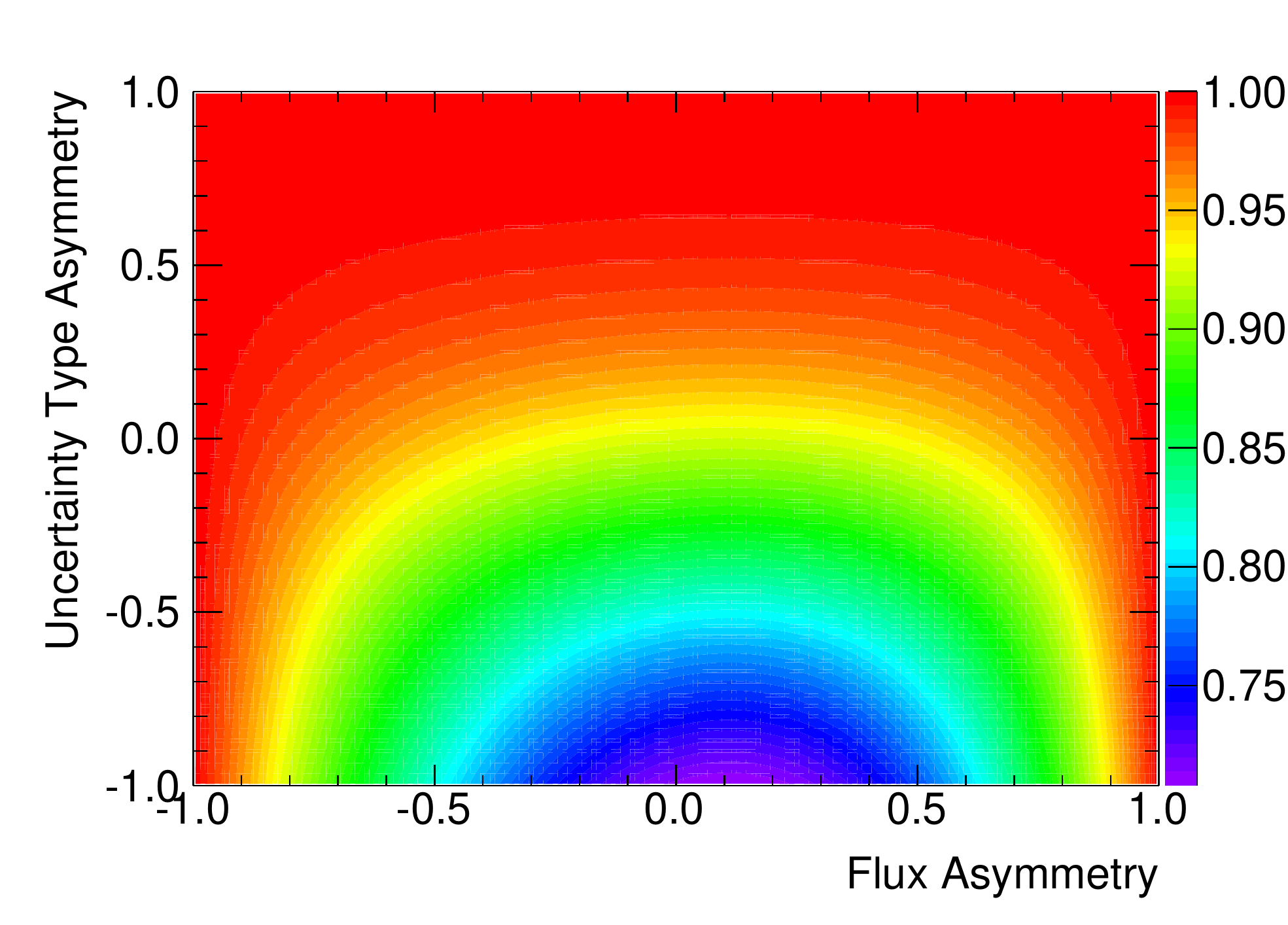}
\caption{ 
	The suppression fraction (SF) for DC-I, coloured coded in the z-axis, is shown against the reactor fluxes asymmetry, defined as $(\Phi_{R2} - \Phi_{R1})/(\Phi_{R2} + \Phi_{R1})$, and the uncertainty correlation asymmetry, defined as $(\delta^c - \delta^u)/(\delta^c + \delta^u)$. 
	No reduction of uncertainties is indicated by an SF=1.0 (red). 
	Whenever one reactor dominates (i.e. the reactor flux asymmetry tends to $\pm 1.0$) or the reactor uncertainties are mostly correlated (i.e. $(\delta^c - \delta^u)/(\delta^c + \delta^u) \to +1.0$), there is small or even no SF. 
	The maximal suppression (violet) is obtained whenever the reactor uncertainties are fully uncorrelated and both reactors are delivering similar fluxes. 
	The expected SF minimum is $1/\sqrt{2}$ ($\sim$0.7), only achievable for a slight offset to R2 to compensate the fact that the far and R1 are closer together, as shown in Fig.~\ref{Fig-Expsetups}. 
	}
\label{Fig-DC_0near_2D}
\end{figure}

The implications of Eq.~\ref{eq_finI1D2R} are best illustrated in Fig.~\ref{Fig-DC_0near_2D} where the SF is shown against the reactor flux asymmetry between both Chooz reactors, defined as $(\Phi_{R2} - \Phi_{R1})/(\Phi_{R2} + \Phi_{R2})$, and the reactor uncertainty type asymmetry, defined as $(\delta^c - \delta^u)/(\delta^c + \delta^u)$, characterising the fraction of reactor error (un)correlation generically. 
There are two interesting limit cases in Fig.~\ref{Fig-DC_0near_2D} where there is no error reduction (i.e. SF~=~1):

\begin{itemize}

\item 
	when the errors are maximally correlated, represented by the condition $(\delta^c - \delta^u)/(\delta^c + \delta^u) \to 1.0$ (or $k \to 1$ in Eq.~\ref{eq_finI1D2R})

\item 
	when either reactor is working, represented by the condition $(\Phi_{R2} - \Phi_{R1})/(\Phi_{R1} + \Phi_{R2}) \to \pm 1.0$

\end{itemize}

\noindent 
this is because those cases cannot be effectively distinguished from the case of {\it single detector with a single reactor} scenario. 
If two reactors uncertainties are totally correlated, this is like if it was one effective reactor, regardless of the geometry of the setup.

Conversely, some error reduction (SF~\textless1) is expected elsewhere: whenever the source is consistent made up of two independent (i.e. uncorrelated errors; i.e. $k = 0$) reactors working at about the same contribution. 
Error suppression is expected to be maximal (i.e. SF minimal) when the two reactors have totally uncorrelated errors, since those reactor errors scale with $1/\sqrt{2}$ ($\sim$0.7).
The slight asymmetry in the reactor flux asymmetry (x-axis) is a consequence of the distance far-R1 being slightly closer than far-R2, as shown in Fig.~\ref{Fig-Expsetups}.
While rather trivial, the one detector case is important for many experiments and it is remarkably illustrative in the context of this discussion, since the fore-mentioned description inverts when considering a multi-detector sites, as described next.

%%%%%%%%%%%%%%%%%%%%%%%%
\subsection{Multi-Detector: the DC-II Configuration}
\label{sec:1nf}
%%%%%%%%%%%%%%%%%%%%%%%%

Let us consider now the SF behaviour for the DC-II scenario with two detectors with two reactors.
In this case, it is more reasonable to characterised the SF associated to the {\it near-far ratio}, most relevant for oscillation analyses, as indicated by

\begin{equation} 
\mathrm{Ratio}(f/n) = \frac{\Phi_f}{\mathrm{C} \times \Phi_n} 
\label{eq:alpha}
\end{equation} 

\noindent when C represents a constant such that the predicted un-oscillated flux at the far is obtained from the near (i.e. $\Phi_f^{pred} = \mathrm{C} \times \Phi_n$). 
This constant accounts for the difference in the baselines between the near and far sites, due to the isotropic emission of neutrinos in reactor.
The near and far are respectively indicated by the subindexes $n$ and $f$.
Let us reconsider Eq.~\ref{eq:defFIgen} now generalised to consider the contribution of the two detectors to compute the ratio $\Phi_f/\Phi_n$. 
As in the case of the single detector case, we are interested in the expression for SF but now relative to the ratio between detectors, indicated by $\delta_{nf}$, to the relative error of the flux emitted by each reactor, thus

\begin{figure*}[ht!]
\centering
\begin{tabular}{cc}
\includegraphics[scale=0.45]{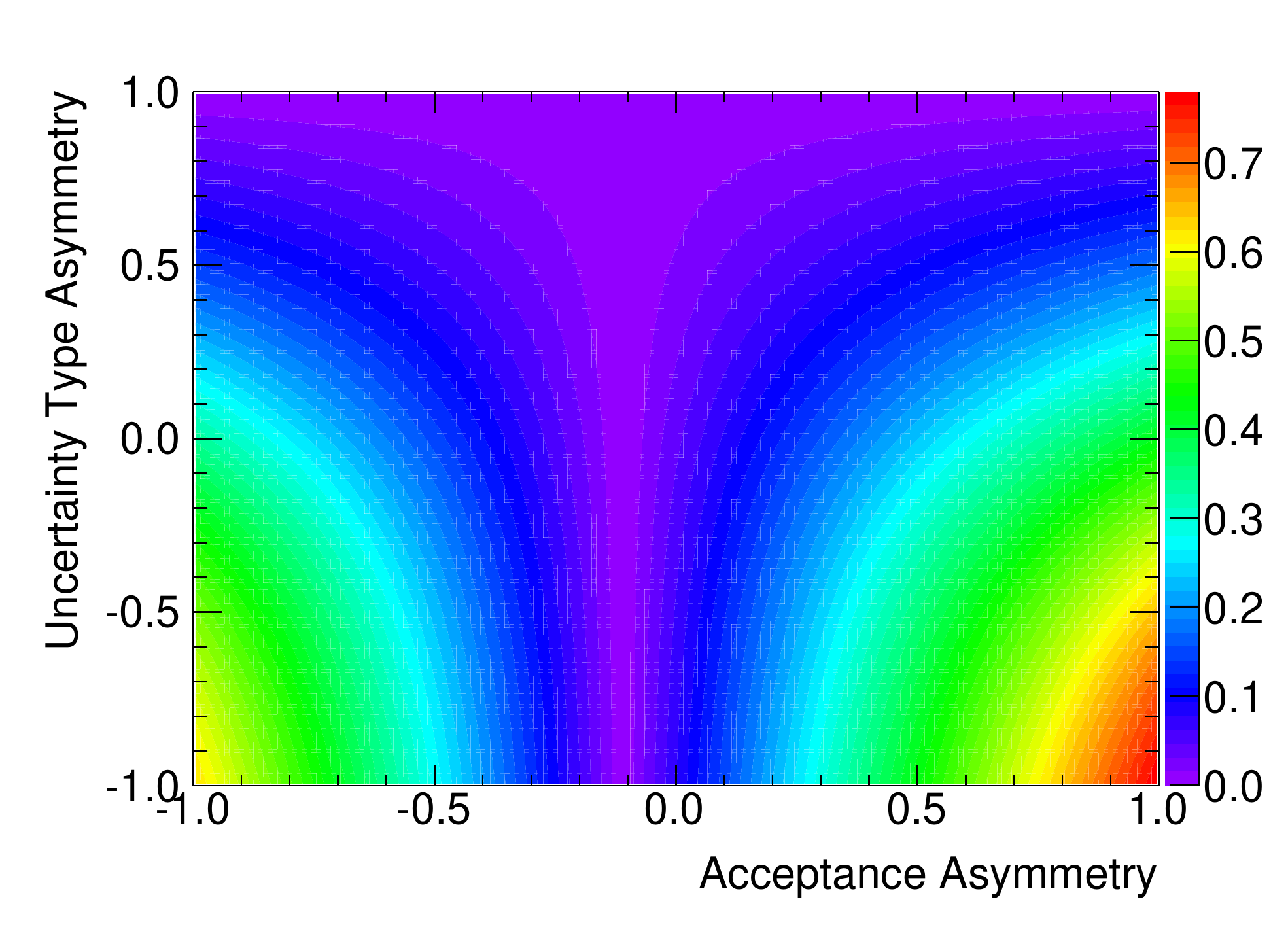} 
&
\includegraphics[scale=0.45]{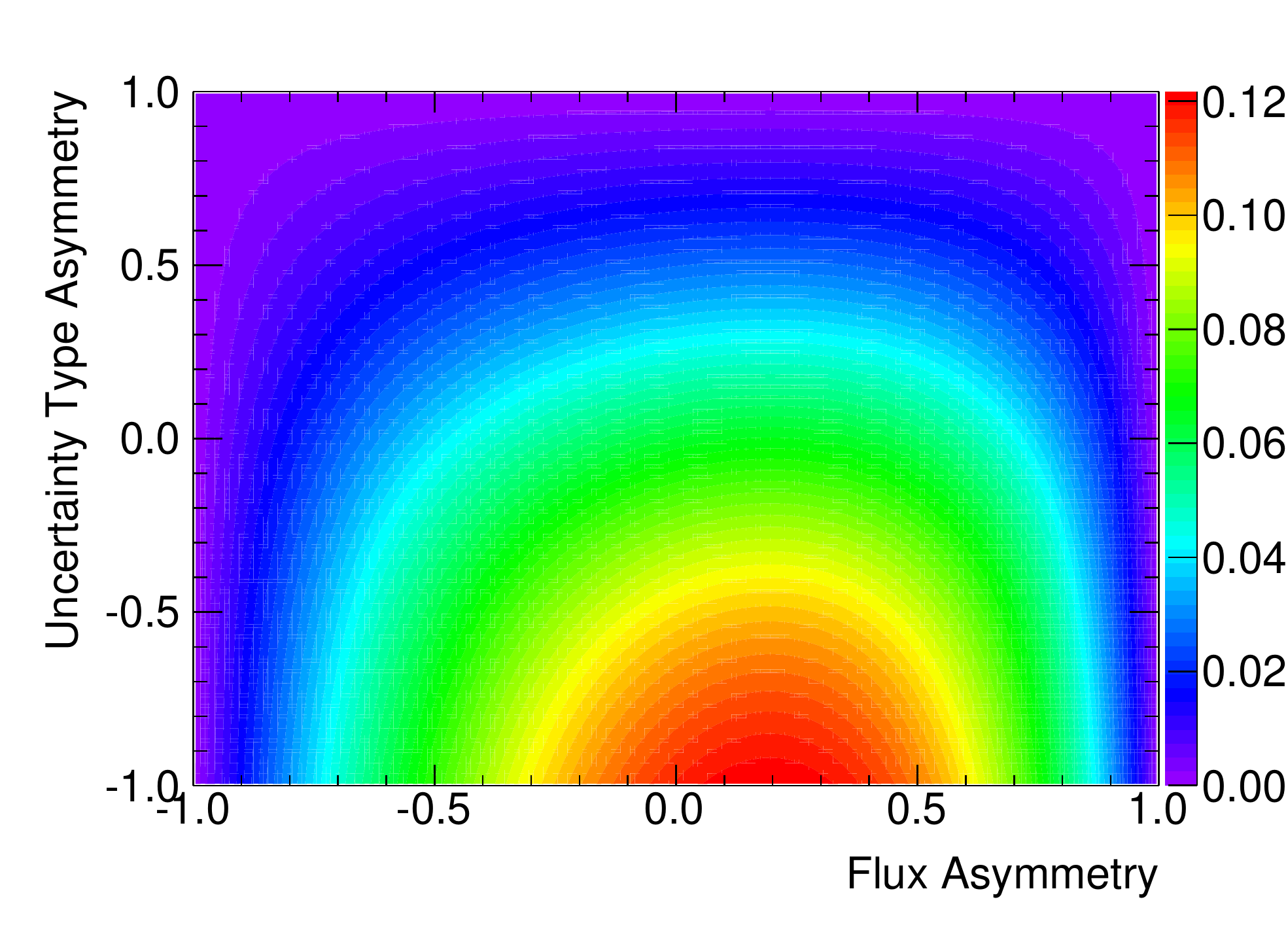}
\end{tabular}
\caption{
	The suppression fraction (SF) for DC-II is illustrated coloured coded in the z-axis. 
	{\bf Left} shows the evolution of SF against both the reactor acceptance asymmetry (x-axis), defined as $(\Omega_{R2}-\Omega_{R1})/(\Omega_{R2}+\Omega_{R1})$, and the reactor uncertainty type asymmetry (y-axis), defined as $(\delta^c-\delta^u)/(\delta^c+\delta^u)$. 
	The iso-flux condition is met whenever the acceptance asymmetry is $\sim$0, leading to a SF~$\to$0 (violet), regardless of the nature of reactor uncertainties, thus demonstrating that the near is a {\it geometrical perfect monitor} of the far. 
	The DC geometry is such that the iso-flux-ness is partially met along the projection $(\Omega_{R2}-\Omega_{R1})/(\Omega_{R2}+\Omega_{R1}) \approx -0.28$, hence SF is, at most, $\sim$0.12.
	{\bf Right} shows the evolution of SF against both the reactor power flux asymmetry (x-axis), defined as $(\Phi_{R2}-\Phi_{R1})/(\Phi_{R2}+\Phi_{R1})$ (i.e. the flux difference between reactor R1 and R2), and the reactor uncertainty type asymmetry (y-axis), as defined before.
	Note that the maximal value possible of SF is the same $\sim$0.12, as explained.
	Total cancellation (i.e. lowest SF) is found if the uncertainties are reactor correlated maximally or either reactor is off. 
	This is because both cases are mathematically identical as to having {\it one effective reactor} as source, perfectly monitored by the near, regardless of the experimental setup geometry.
	}
\label{Fig-DC_1near_2D}
\end{figure*}

\begin{equation} 
\mathrm{SF}^2 = \left( \frac{ \delta_{nf} }{\delta_R} \right)^2 = \left( \frac{\delta_n}{\delta_R}\right)^2 + \left( \frac{\delta_f}{\delta_R}\right)^2 - 2 \, \frac{\mathrm{cov}(\Phi_n, \Phi_f)}{\delta_R^2 \Phi_n \Phi_f} 
\label{eq:I2D2Rintro}
\end{equation} 

\noindent 
where $\delta_n$ and $\delta_f$ are the relative errors of the fluxes reaching the individual detectors, respectively for the near and far. 
The first two terms in Eq.~\ref{eq:I2D2Rintro} are the same as the ones obtained for single detector configuration (Eq.~\ref{eq_finI1D2R}). 
The new term corresponds to the covariant term computed as

\begin{eqnarray}
\mathrm{cov}(\Phi_n, \Phi_f) &=& \delta_R^2 [ \Omega_{nR1} \Omega_{fR1} \Phi_{R1}^2 \nonumber\\
&& +~\Omega_{nR2} \Omega_{fR2} \Phi_{R2}^2\nonumber\\
&& +~k \, \Phi_{R1} \Phi_{R2} ( \Omega_{nR1} \Omega_{fR2}\nonumber\\
&& +~\Omega_{nR2} \Omega_{fR1} ) ]
\label{eq_covdef}
\end{eqnarray}

\noindent 
where k has the same definition as in Eq.~\ref{eq:kdef} for the error correlation factor. 
We introduce $\Omega_{nR}$ and $\Omega_{fR}$ parameters, as in Eq.~\ref{eq:defOMR}, to obtain now

\begin{equation}
\frac{\mathrm{cov}(\Phi_n, \Phi_f)}{\delta_R^2 \Phi_n \Phi_f} = \frac{1 + \Omega_{nR}\Omega_{fR} + k \, (\Omega_{nR} + \Omega_{fR})}{(1 + \Omega_{fR})( 1 + \Omega_{nR} )} 
\end{equation}

\noindent
such that the previous Eq.~\ref{eq:I2D2Rintro} now becomes

\begin{eqnarray}
\mathrm{SF}^2 &=& \frac{ 1 + \Omega_{nR}^2 + 2 k \Omega_{nR}}{ (1 + \Omega_{nR})^2 } \nonumber \\
&& +~\frac{ 1 + \Omega_{fR}^2 + 2 k \Omega_{fR}}{ (1 + \Omega_{fR})^2 } \nonumber \\ 
&& -~2 \, \frac{1 + \Omega_{nR}\Omega_{fR} + k(\Omega_{nR} + \Omega_{fR})}{(1 + \Omega_{fR})( 1 + \Omega_{nR} )}
\label{eq_I2D2R_final}
\end{eqnarray}

\noindent 
representing SF general expression for the near-far setup with two reactors.

As before, $\Omega_{nR}$ and $\Omega_{fR}$ are independent from any factor which multiplies both solid angles or both reactor rates, thus the weight parameter from Eq.~\ref{eq:alpha} does not contribute to the expression from Eq.~\ref{eq_I2D2R_final}. 
This expression is also independent from the transformation $\Omega_{DR} \to 1/\Omega_{DR}$ (with D = n,f) given by a different way of defining the fraction terms in Eq.~\ref{eq:defOMR}. 
By definition, the iso-flux condition implies that the relative antineutrino fluxes from reactors is the same for both detectors. This can be mathematically represented as

\begin{equation}
  \Omega_{nR} = \Omega_{fR}
 \label{isofl_nf}
\end{equation}

The implication of Eq.~\ref{eq_I2D2R_final} are best illustrated in Fig.~\ref{Fig-DC_1near_2D} where the evolution of the SF is shown. 
Fig.~\ref{Fig-DC_1near_2D}-Left plot shows SF evolution relative to both the reactor acceptance asymmetry, defined as $(\Omega_{R2}-\Omega_{R1})/(\Omega_{R2}+\Omega_{R1})$, and the reactor uncertainty type asymmetry, defined as $(\delta^c-\delta^u)/(\delta^c+\delta^u)$. 
The iso-flux condition is met whenever the reactor acceptance asymmetry is $\sim$0, leading to a SF to fully cancel, regardless of the nature of reactor uncertainties, demonstrating so the near is a {\it geometrical perfect monitor} to the far. 
This is the same as imposing the condition in Eq.~\ref{isofl_nf} in Eq.~\ref{eq_I2D2R_final}.
The DC geometry is such that the iso-flux-ness is partially met along the projection $(\Omega_{R2}-\Omega_{R1})/(\Omega_{R2}+\Omega_{R1}) \approx -0.28$, hence a major reduction of SF is obtained where SF can, at most, be $\sim$0.12; i.e. $\sim$90\% of the original uncertainty is thus suppressed. 
Fig.~\ref{Fig-DC_1near_2D}-Right plot shows SF evolution relative to both the reactor power flux asymmetry, defined as $(\Phi_{R2}-\Phi_{R1})/(\Phi_{R2}+\Phi_{R1})$, and the reactor uncertainty type asymmetry, as defined before.
The effect of the reactor flux asymmetry can only exemplified in the case of DC with two reactors.
Total cancellation of SF also is found if the uncertainties are reactor correlated maximally (i.e. $k = 1$) or either reactor is off. 
This is expected because both such conditions are equivalent as having {\it one effective reactor} as source, thus perfectly monitored by the near, regardless of the experimental setup geometry or the type of the uncertainty.
As expected, this conclusion is general and true independently of how many reactors are considered.
This implies that, for example, whenever DC runs with 1 reactor and both detectors, the flux error is zero for that data set.
It is worth noting that the pattern shown for DC-II, illustrated in Fig.~\ref{Fig-DC_1near_2D}-Right, is exactly the opposite to the one exhibited by DC-I, illustrated in Fig.~\ref{Fig-DC_0near_2D}, where no suppression (SF=1) is obtained in the case of total correlated reactor uncertainties.
In brief, the DC-II performance, SF~$\sim$0.12, exemplifies the best multi-detector error cancellation due its almost iso-flux geometry where almost $\sim$90\% of the SF is due geometry.
The remaining SF fraction ($\sim$30\%) is caused by the $1/\sqrt{N_R}$ scaling of the remaining totally uncorrelated error, where $N_R = 2$.

%Finally, it is worth noting that most of the conclusions here described are applicable to any isotropic neutrino sources, such as the reactors, but not only.
%Conversely, when considering non-isotropic sources, such as popular meson-decay-based neutrino beam multi-detector experiments, the flux systematics cancellation between near and far is impractical and delicate extrapolation methods are needed, including the cost of systematics.

%%%%%%%%%%%%%%%%%%%%%%%%
\subsection{Multi-Detector: the RENO Configuration}
\label{sec:1nf_re}
%%%%%%%%%%%%%%%%%%%%%%%%

Generalising Eq.~\ref{eq_I2D2R_final} to account for a larger number of reactors $(N_R = 6)$, RENO is becomes our next case with two detector sites and $N_R = 6$, as illustrated in Fig.~\ref{Fig-Expsetups}. 
The new general expression is

\begin{eqnarray}
\mathrm{SF}^2  &=& \sum_{d = n,f} \left( \frac{ 1 + S_{d^2} + 2 k ( S_d + S'_d ) }{ (1 + S_d)^2 } \right) \nonumber \\
&& -~2 \, \frac{1 + S_{nf} + k(S_n + S_f + S'_{nf})}{(1 + S_n)( 1 + S_f)}
\label{eq_I2D2nR_final}
\end{eqnarray}

\noindent where we have introduced the following notation

\begin{eqnarray}
&& S_d = \sum_{i>1}^{N_R-1} \Omega^{(i)}_{dR}; \nonumber \\ 
&& S_{d^2} = \sum_{i>1}^{N_R-1} (\Omega^{(i)}_{dR})^2; \nonumber \\ 
&& S'_d = \sum_{i,j>1;i \neq j}^{N_R-1} \Omega^{(i)}_{dR} \Omega^{(j)}_{dR}; \nonumber \\
&& S_{nf} = \sum_{i>1}^{N_R-1} \Omega^{(i)}_{nR} \Omega^{(i)}_{fR}; \nonumber \\ 
&& S'_{nf} = \sum_{i,j>1;i \neq j}^{N_R-1} \Omega^{(i)}_{nR} \Omega^{(j)}_{fR}; \nonumber 
\label{eq:sumsparams}
\end{eqnarray}

\noindent and $\Omega^{(i)}_{dR}$ is a generalization from the Eq.~\ref{eq:defOMR}, to obtain

\begin{equation} 
\Omega^{(i)}_{dR} = \frac{\Omega_{dRi} \Phi_{Ri}}{\Omega_{dR1} \Phi_{R1}} \cong \frac{L_{dR1}^2 \Phi_{Ri}}{L_{dRi}^2 \Phi_{R1}} ; \quad i = 2 \dots N_R-1     
\label{eq:defOMRi}
\end{equation} 

\noindent where reactor R1 has been arbitrary chosen as reference reactor. Nevertheless, choosing a different reactor leave Eq.~\ref{eq_I2D2nR_final} unchanged. 

The iso-flux condition, originally in Eq.~\ref{isofl_nf}, is now more generally expressed as

\begin{equation} 
\Omega^{(i)}_{nR} = \Omega^{(i)}_{fR}
\label{isofl_nf_gener}
\end{equation} 

\noindent which, like in the case of DC-II, would cause SF to go to zero, regardless of any other error dependence, upon imposition, if the geometry of the experimental site considered allows.

Upon introducing the geometry of RENO, we can summarise our finding as follows,

\begin{itemize}

\item 
	SF will obtained full cancellation if reactor errors are fully correlated, as demonstrated and illustrated for DC-II (Fig.~\ref{Fig-DC_1near_2D}), regardless of the experimental site geometry.

\item 
	SF will be largest in case of the fully uncorrelated reactor systematics errors when they deliver similar fluxes, also as obtained for DC-II. 		However, an overall SF follows $1/\sqrt{N_R}$ reduction, hence, SF will benefit sites with many independent (i.e. uncorrelated errors) identical reactors. 
	In a sites with different reactor types, the effective SF might be deteriorated.

	Mathematically, this can be seen when considering $k = 0$ and $N_R \to \infty$ into Eq.~\ref{eq_I2D2nR_final}. 
	The factors on the denominator, $(1 + S_d)^2$ and $(1 + S_n)(1 + S_f)$ will increase faster than the corresponding factors at numerator, $S_{d^2}$ respectively $S_{nf}$, hence leading to cancellation.

\item 
	RENO has some SF reduction due to some degree of iso-flux-ness. 
	The amount of SF reduction is limited, as expected, since the geometry of the site is not optimal to ensure that the near and far have a similar 	contribution across all reactors equally.
	The near sees, respectively, $\sim$78\%, $\sim$16\% and $\sim$6\% from the different reactors pairs (from the nearest to the farthest), while the 	far sees a more even contribution from all reactors. 
	Thus, the RENO site is not expected to meet the iso-flux condition fully.
	The total SF measured was estimated to be $\sim$0.23 in the full power scenario.
%Out of this SF obtained, only $\sim$19 is due to partial iso-flux condition.

\item
	There is a small difference between the nominal power of two RENO reactors~\cite{Ahn:2012nd}. 
	In our analysis, we took this into account by weighting the correspondent $\Omega^{(i)}_{DR}$ terms. 

\end{itemize}

Unlike so far assumed, commercial reactors do not deliver their flux constantly over time.
Some unavoidable variations expected are due to refuelling, a total reactor stop once a year over a few weeks, and reactor burn-up effects, typically a few percent decrease in flux over the entire year.
In addition, reactors might be run differently every year, so each reactor cycle history is a priori expected to be unique.
When that happens in site with many reactors, the effective SF can only deteriorate as the two above conditions will be varied away from its optimal spot.
This effect will be further studied and quantified later on.

%%%%%%%%%%%%%%%%%%%%%%%%
\subsection{Multi-Detector: the Daya Bay Configuration}
%%%%%%%%%%%%%%%%%%%%%%%%

Let us now consider the setup of the DB experiment now. 
As illustrated in Fig.~\ref{Fig-Expsetups}, this site differs from the other on having 2 near sites to monitor the two sets of reactors geometrically grouped of the power plant. 
The near-1 monitors mainly reactors R1 and R2, while and the near-2 monitors all others.

Using similar formalism till now, the only modification here is to build the Ratio($f/n$), used in Eq.~\ref{eq:alpha}, with the explicit contribution of two near sites in the denominator. Hence, the new expression becomes

\begin{equation} 
\mathrm{Ratio}(f/n) = \frac{\Phi_f}{\beta \Phi_{n1} + \gamma \, \Phi_{n2}} 
\label{eq:gammarho}
\end{equation} 

where $\beta$ and $\gamma$ are two constants, whose values are fixed by the experimental geometry and the fluxes of the running reactors. 
If optimisation of the site, upon design, wanted to be considered, the effective SF of DB would have depended on them. 
For the prediction, the new flux is thus constructed from the combination of the two near sites measurements; i.e. $\Phi_{n1}$, $\Phi_{n2}$. 
However, the near sites do not only monitor their respective reactors, but instead, they also see a fraction of the other reactors (not supposed to be monitoring); i.e. near-2 also sees $\sim$6.5\% of reactors R1 and R2 and near-1 see about, $\sim$17\% of the remaining reactors.
This makes this ratio very delicate, since some degree of double-counting is unavoidable by the nears is unavoidable, while this is not present in the far. 
This will translate into a loss of the iso-flux-ness condition.
The same conclusions listed for RENO are valid for DB (previous section).
The overall SF estimated for this site is $0.18$ benefiting for slightly better partial iso-flux-ness relative to RENO, despite the visual appealing geometrical symmetry of the RENO site.
In the case of both RENO and DB, most of the SF ($\sim$60\%) arises from the $1/\sqrt{N_R}$ scaling due to their large number of identical reactors.

%%%%%%%%%%%%%%%%%%%%%%%%
\subsection{Suppression Fraction Estimation via Monte-Carlo}
%%%%%%%%%%%%%%%%%%%%%%%%

Together with the analytical formalism presented, we have analysed the reactor flux systematics using a Monte-Carlo based flux simulation method. Likewise, the full geometry and ingredients to each experimental sites are simulated.
Each reactor is described by its thermal power (P$_{th}$) and fractional fission rates ($\alpha_k$ with $k = ^{235}$U, $^{238}$U, $^{239}$Pu and $^{241}$Pu, the main core isotopes), while each detector is described by the number of free protons in the target and by the detection efficiency.
For simplicity, the same mean $\alpha_k$'s have been used for all experiments, however this is not totally correct, as different experiments not only have different reactors but also the fuel composition, namely the enrichment of $^{235}$U, is expected to be somewhat different.
%In the followings, the uncertainties of the reactor parameters (P$_{th}$ and $\alpha_k$) are assumed to be measured fully uncorrelated by detectors. %while the ones of physics-related parameters such as the energy release per fission and the number of inverse beta decays per fission are fully correlated, thus cancelling . 
Since the detector correlated errors are canceled in multi-detector setups, we have propagated only the detector uncorrelated errors.

The antineutrino flux emitted by a given reactor is a function of thermal power and the fission rates define 

\begin{equation}
\Phi = \Phi (P_{th}, \alpha_k).
\end{equation}

\noindent As the error propagation formalism does not depend on the specific form of the function $\Phi$, we do not describe it in detail. For a given binning of the antineutrino energy spectrum, the error of the flux is expressed as the covariance matrix

\begin{equation}
M^\Phi_{ij} = M^{P_{th}}_{ij} + M^{\alpha}_{ij},   
\end{equation}

\noindent where $i$ and $j$ stand for the energy bins, and $M^{P_{th}}_{ij}\equiv \langle \delta_{P_{th}}^i \; \delta_{P_{th}}^j \rangle$ and $M_{ij}^{\alpha} \equiv \langle \delta_{\alpha}^i \; \delta_{\alpha}^j \rangle$ are the uncertainty contributions from the thermal power and the fission rates, respectively. As the error on P$_{th}$ does not depend on the antineutrino energy, $M^{P_{th}}$ is a diagonal matrix with $\delta_{P_{th}}^i=\delta_{P_{th}}^j=\delta_{P_{th}}$.

For a given reactor $r$, the error on $\Phi$ due to the thermal power uncertainty $\delta_{P_{th}}$ is computed as

\begin{equation}
\label{eq:ePth}
\delta_{P_{th}}^r = \Phi_r(P_{th}^r + \sigma_{P_{th}}^r, \alpha_k) - \Phi_r(P_{th}, \alpha_k),  
\end{equation}

\noindent where $\sigma_{P_{th}}^r$ is the 1-$\sigma$ error of the thermal power. 

$M_{ij}^{\alpha}$ is computed by propagating the covariance matrix of the fission rates, $C_{kl}^\alpha \equiv \rho_{kl} \sigma_{\alpha_k} \sigma_{\alpha_l}$, being $\rho$ the correlation matrix and $\sigma_{\alpha}$ the 1-$\sigma$ error of the fission rate

\begin{equation}
M_{ij}^{\alpha,r} = \sum_{k,l}^{4} D_{ik}C_{kl}^\alpha D_{jl}^T,
\end{equation}

\noindent where $D$ stands for the derivatives defined as

\begin{equation}
D_{ik} \equiv \frac{ \Phi_r^i(\alpha_k + \sigma_k, P_{th}) - \Phi_r^i(\alpha_k, P_{th})}{\sigma_k}.
\end{equation}

The $\alpha$-related error on the integrated flux from a reactor $R$ can be expressed as

\begin{equation}
\label{eq:eAlpha}
\delta_{\alpha}^R = \sqrt{\sum_{ij} E_i M_{ij}^{\alpha,R} E_j},
\end{equation}

\noindent where $E_i$ stands for the energy phase space of bin $i$.

For an experimental setup with $N_R$ reactors, the final integrated error on the antineutrino flux  is derived as the quadratic sum of the contributions from individual reactors, provided by Eq.~\ref{eq:ePth} and Eq.~\ref{eq:eAlpha}

\begin{equation}
\delta_{\Phi} = \sqrt{\sum_{r}^{N_{R}} (\delta_{P_{th}}^r)^2 + (\delta_{\alpha}^r)^2 }.
\end{equation}

Using the procedure described above, we computed the central values ($\Phi^n$ and $\Phi^f$) and uncertainties ($\delta^n$ and $\delta^f$) of the fluxes at both near ($n$) and far ($f$) detectors of DC-II, DB and RENO experiments. 
We also computed the far-to-near flux ratio, ratio($fn$) = $\Phi^f/\Phi^n$, along with its error and, finaly, derived the error SF as

\begin{equation}
SF = \frac{\delta_{NF}^{f}}{\delta_{f}},
\end{equation}

\noindent where $\delta_{NF}^{f}$ is the error on the flux at the far detector when computed as $\mathrm{ratio}(fn)\times \Phi^n$. Such a SF is independent of the parameterisation of $\Phi$ and the specific values of $\sigma_{P_{th}}$ and $\sigma_{\alpha}$.
% and $\rho$, which therefore have been set arbitrarily for this study. LET'S NOT TALK MORE ABOUT RHO, PLEASE.

In this calculation, the SF accounts only for the error suppression provided by the iso-flux-ness of the experimental setup. 
In order to be compared with the analytical estimation, developed in the previous sections based on the $\delta^f/\delta_R$ ratio, one needs to incorporate the uncorrelated error reduction due to the number of reactors. 
If detectors fulfilled the iso-flux condition, such a reduction would be $\sqrt{N_R}$. 
As this is not the case for any of the experiments considered, an effective number of reactors ($N_{R}^{eff}$) has to be computed. 
According to our simulation, the $N_{R}^{eff}$ values are 1.98, 5.95 and 5.78, for DC, DB and RENO, respectively.

The SF obtained via Monte-Carlo, once corrected by the $N_{R}^{eff}$, are in remarkably good agreement with the ones from the analytic calculations:
identical numbers reproduced to the 3$^{rd}$ digit.
This provides a mutual validation of the procedures presented.
% and the hypotheses made to obtain both estimations. 
%[MURIEL: here say which hypotheses have been validated] 
%The final results are listed in Table~\ref{Tab-allcomp}.

%%%%%%%%%%%%%%%%%%%%%%%%
\subsection{Reactor Time Variations: Refuelling Scenario}
%%%%%%%%%%%%%%%%%%%%%%%%

Considering time variation of the flux delivered per reactor will modify the effective SF behaviour described so far, as the effective site reactor power configuration departs from the full-power over-simplified scenario considered so far.
In this section, we shall consider the impact to the effective SF for each experiments upon reactor refuelling estimated analytically, which a more realistic experimental scenario since this is, by far, the largest expected reactor-to-reactor variation that can be modelled similarly to all sites considered.
The effects associated to burn-up, fuel composition, variations caused by running operation constraints (unicity reactor cycle) will all be neglected, as they are expected to be both smaller in magnitude and site dependent, thus out of the scope of the generic description considered here.
However, more accurate estimations expected by each specific experiment should consider its impact carefully.
Therefore, for simplicity, let us consider the following refuelling scenario: for a total period of one year, each reactor is stopped for two months for refuelling, while during this period the other reactors are running.
In this scenario, we consider also fully uncorrelated reactors errors, representing the worst case scenario. 
The impact on SF due to one reactor going down is, obviously, site dependent. 
For DC-II, the overall SF is zero whenever one reactor is monitored by two detectors, regardless of the geometry, as demonstrated before.
For RENO and DB, contrary to DC-II, we expect an effective deterioration of SF since both the effective iso-flux-ness condition (power symmetry geometry) and the number of reactors running worsens upon refuelling. 

Our results can be summarised as follows:

\begin{description}

\item[DC-II:] 
	the year average SF is 0.08 (refuelling scenario), to be compared with 0.12 (full power).
	Any one reactor off periods will only benefit DC-II as it has null flux uncertainty, therefore the longer the better for systematics, although that will imply a loss in luminosity.

\item[DB:] 
	the tear average SF is 0.20, against the 0.18 obtained at full power.

\item[RENO:] 
	the results are shown in Table~\ref{Tab-ReSc} for each reactor, however the year average SF was found to be 0.24, to be compared when considering all reactors (0.23).

\end{description}

\begin{table}
\caption{
	SF values for different RENO reactor configurations: all ON and each off.
	Notation as shown in Fig.~\ref{Fig-Expsetups}.
	}
\begin{ruledtabular}
\begin{tabular}{ccccccc}
ON & R1$^{\mathrm{OFF}}$ & R2$^{\mathrm{OFF}}$ & R3$^{\mathrm{OFF}}$ & R4$^{\mathrm{OFF}}$ & R5$^{\mathrm{OFF}}$ & R6$^{\mathrm{OFF}}$ \\
\hline
0.227 & 0.226 & 0.265 & 0.234 & 0.285 & 0.244 & 0.206
\end{tabular}
\end{ruledtabular}
\label{Tab-ReSc}
\end{table}

%%%%0:  0.226898
%%%%1:  0.226228
%%%%2:  0.26447
%%%%3:  0.234038
%%%%4:  0.284841
%%%%5:  0.244214
%%%%6:  0.205937

Finally, the refuelling scenario here described (one reactor stopped at the time) is specially naive in the case of RENO and DB, as their data shows simultaneous stop of up 3 reactors for some time, hence the site power might swing up to $\sim$50\%.
This will have implications, typically, towards the deterioration of the effective SF to be computed for those sites, when integrating over all those effects, including fuel burn-up.
That level of accuracy is beyond the scope of this estimation.

%%%%%%%%%%%%%%%%%%%%%%%%
\subsection{Daya Bay Near Detectors Optimisation}
%%%%%%%%%%%%%%%%%%%%%%%%

As indicated in Eq.~\ref{eq:gammarho}, combining flux measurements from different nears with different weights seems an appealing consequence for the case of DB. 
This represents a fundamental difference with respect to the experimental configuration with a single near.
Any scaling of a flux measurement in Eq.~\ref{eq:defFIgen} is equivalent to a change to the baseline of the detectors.
So, a difference between the values of $\beta$ and $\gamma$, in Eq.~\ref{eq:gammarho}, is equivalent to a change between the relative baselines of the near sites.
The authors think that such transformation might become problematic given two possible effects:

\begin{figure}[htb!]
\centering
\includegraphics[scale=0.45]{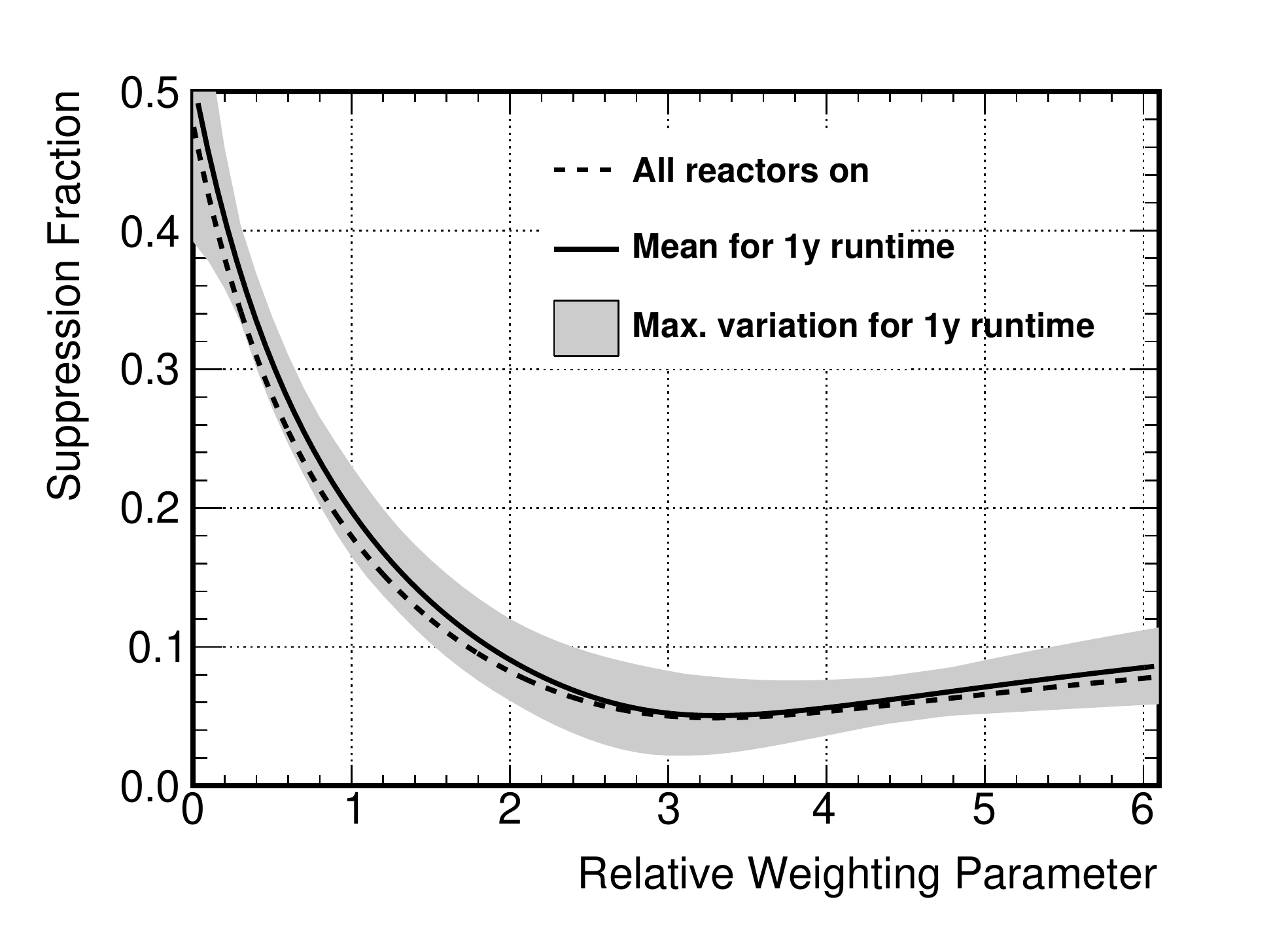}
\caption{
	The SF variation of DB for the near sites relative weighting is illustrated. 
	The minimum obtained is in agreement with DB published results.
	The case for all reactors at full power is indicated by the dashed line, while the maximal variations obtained due to refuelling scenario (simplified case described in text) is indicated by the shaded region.
	The physical interpretation of this analysis and its application for meaningful reactor flux error suppression is a debate with no consensus within the field.
	Our result is here illustrated as validation of published results, but it will disregarded for further discussion.
	}
\label{Fig-DB_tune}
\end{figure}

\begin{itemize}

\item 
	the near sites do not only monitor the closest reactors, but there is a sizeable fraction of flux coming from the further reactors.
	This implies that the relative normalisation of each near are not fully independent.

\item 
	the effect of the neutrino oscillations measured by the near detectors are not fully negligible.

\end{itemize}

If we ignored both physical conditions, we could consider the optimisation of the contribution of the near sites, as done by DB in~\cite{DB2}.
As we have shown previously, our formalism is transparent to any factor which multiplies all solid angles or all fluxes provided by reactors with the same factor for a given detector. 
Thus, we consider only the relative difference of the weighting parameters as

\begin{equation}
\Phi_n \equiv \Phi^{prediction}_f = \Phi_{n_1} + \rho \, \Phi_{n_2} = \sum_{i}^{N_R} \Omega_{nRi}(\rho) \Phi_{Ri}
\label{eq:defPhi_n1n2}
\end{equation}

\noindent where $\rho = \gamma/\beta$ is the relative weight parameter and

\begin{equation}
\Omega_{nRi}(\rho) = \Omega_{n_1Ri} + \rho \, \Omega_{n_2Ri}
\label{eq:defOni_n1n2}
\end{equation}

\noindent is a linear combination of the omega terms as described by Eq.~\ref{eq:defOMRi}.

\begin{table*}
\caption{
	The analytical estimated {\it suppression fraction} (SF) per experiment are here summarised. 
	The SF estimated via MC are not shown, as they are numerical identical up to the third digit.
	SF(total) is shown in both the full-power (over-simplified) and the refuelling scenarios.
	The SF(total) is defined as SF(total)~$=$~SF(iso-flux)~$\times$~SF($N_R$)~$\times$~SF(correlation), whose value is within the interval [0,1], implying total and no suppression, respectively.
	%broken down into its components SF(iso-flux) and SF($N_R$), such that SF(total)~$=$~SF(iso-flux)$~\times$~SF($N_R$).
	The potential exploitation of SF(correlation) suppression needs dedicated analysis beyond the scope of this publication.
	Hence the remaining error is assumed to be totally reactor uncorrelated, implying a SF(correlation)=1 (i.e. no suppression), being the most conservative scenario for multi-detector and multi-reactor experiments.
	Note, however, that this implies maximal suppression due to SF($N_R$).
	The impact of SF(correlation), while not estimated for any specific case (being strongly reactor dependent) is generically illustrated in Fig.~\ref{Fig-ALL_1D}.
	%Rough breakdown of SF is provided into the iso-flux component and the number of reactor (i.e. $1/\sqrt{N_R}$ of the uncorrelated contribution) as guideline to better appreciate the origin of the SF quoted per site.
	%The total and effective error are quoted in \%, while the SF are quoted in fraction. 
	%The effective error is quantified as the total error (see Table~\ref{Tab-uncorpar}) reduced by its corresponding SF. 
	%ected, since further SF profiting reactor error correlations is unsettled topic in the field worth dedicated debate (out of focus this publication).
	% , expected to reduced, upon error propagation by the $1/\sqrt{N_R}$, where $N_R$ is the number of reactors involved. 
	%The expected $1/\sqrt{N_R}$ reduction is also shown, since the difference relative to the quoted SF indicates the effective contribution due to iso-flux-ness of each site.
	}
\begin{ruledtabular}
\begin{tabular}{l c c c c c}
Experiment     & $N_{R}$ (via MC)   & SF(full-power)            & SF(refuelling)           & SF(iso-flux)   & SF($N_R$)\\
%               &                   & ({\it full-power})   & ({\it refuelling})   &                & \\ 
\hline
Daya Bay       & 6 ($\sim$6.0)      & 0.18                 & 0.20                 & 0.49           & 0.41\\
Double Chooz   & 2 ($\sim$2.0)      & 0.12                 & 0.08                 & 0.11           & 0.71\\
RENO           & 6 ($\sim$5.8)      & 0.23                 & 0.24                 & 0.59           & 0.41\\
\end{tabular}
\end{ruledtabular}
\label{Tab-SF}
\end{table*}

Since both $\Phi_n$ and $\Phi_f$ are given by Eq.~\ref{eq:defFIgen}, we can use directly Eq.~\ref{eq_I2D2nR_final} in order to get the expression for the corresponding SF. 
Inserting the values of the geometrical parameters, we show the variation of the SF relative weighting parameter in Fig.~\ref{Fig-DB_tune} for fully uncorrelated errors of the reactor fluxes. 
The dotted curve represents the variation of the SF when all reactors are operational and the continuous curve represents the mean value between the values obtained when one reactor is off upon reactor refuelling scenario considered in the last section. 
The variations in such refuelling scenario are indicated by the grey area in Fig.~\ref{Fig-DB_tune} accounting for the different reactors.
There are two interesting cases to be highlighted

\begin{description}

\item[$\rho$ = 1:] this point represents no relative weighting applied on the near sites, for which SF is $\sim$0.20.
The spread of SF when considering the reactor refuelling scenario is $\sim$2\%.

\item[$\rho$ at minimum:] the effective SF obtained is about $\sim$0.05. 
The spread due to refuelling is about $\sim$6\%.

\end{description}

Our result is consistent with the official result of DB~\cite{DB2}, where they obtained $\beta$ = 0.04 and $\gamma$ = 0.3 for the analysed data period, considering that one of the near contains twice the events than the other, we obtained $\rho$ to be 3.38.
Using MC, we also obtained respectively, 0.04 and 0.3.
Our calculation here, thus serves as a replication. 
However, due to the difficulties on the physical interpretation of this, the corresponding SF is ignored any further in this publication.
This is consistent with the fact that DB discards it for the measurement of $\theta_{13}$.

%%%%%%%%%%%%%%%%%%%%%%%%
\section{Summary \& Discussion}
%%%%%%%%%%%%%%%%%%%%%%%%

%However, there is no treatment consensus across the collaborations and non-negligible obscurity remains on the origin of the numbers quoted, since there is very little (or nothing) published by the collaborations.
%This lack of literature in the topic opposes the critical role played by the reactor flux systematics, even on multi-detector setups.
%Therefore, we strongly encourage all the collaborations to improve their publication strategy on this critical topic, relevant on the final error of $\theta_{13}$.

We have identified, studied and quantified three mechanisms inducing reactor flux uncertainty suppression in the context of multi-detector experiments in multi-reactor sites.
We have quantified the integral error suppression by the {\it suppression fraction} (SF) analytically (cross-checked via MC) using simplified experimental scenarios to allow coherent relative comparison across experiments.
SF can take values within [0,1], where the extreme cases stand, respectively, for total suppression (SF=0) and no suppression(SF=1).
The three mechanisms can be characterised by their respective SF terms, since SF(total) is defined as SF(total)~$=$~SF($N_R$)~$\times$~SF(iso-flux)~$\times$~SF(correlation), where 
{\it i}) SF($N_R$) is linked to $1/\sqrt{N_R}$ scaling of the remaining uncorrelated reactor error,
{\it ii}) SF(iso-flux) is linked to the site iso-flux condition
and {\it iii}) SF(correlation) is linked to the nature of reactor errors.
Both SF(iso-flux) and SF(correlation) could lead to total error suppression under specific site conditions.
If those terms are to be exploited, those conditions must be carefully evaluated and demonstrated by each experiments accounting accurately for all pertinent effects, although no experiment have ever done this.
Our final results relying simplified refuelling scenarios are summarised in Table~\ref{Tab-SF}.
This results are not expected to be used by experiments, but as mere guidelines for more accurate estimations to follow up.
The contribution of SF(correlation) was not quantified for any specific experimental setup, as it deserves more careful treatment discussed below.
%As guideline, an indicative final flux error could be obtained by combining the total flux error per reactor, summarised in Table~\ref{Tab-uncorpar}, and the corresponding SF(total), summarised in Table~\ref{Tab-SF}, noting that the DC-II error quoted was inferred unofficially from the latest DC-I publication.

The SF($N_R$) has been actively exploited in publications by experiments to improve our knowledge on $\theta_{13}$ under the assumption that the remaining error is is fully reactor uncorrelated.
Typically, $N_R$ refers to the number of effectively identical reactors per site.
%In the case, non-identical reactors, the error should be estimate  
In the case of DB and RENO, this term amounts to $\sim$60\% suppression (6 reactors), whereas this yields only $\sim$30\% suppression for DC-II.
Typically, SF($N_R$) is propagated into the $\theta_{13}$ precision as a byproduct of the $\chi^2$ minimisation.

The estimation of SF(iso-flux) has not been applied by any experiment so far.
It is likely impractically to be implemented via the $\chi^2$ minimisation formulation, instead calculations might follows the prescription here presented.
%, as it demands detailed accounting each reactor-detector pair acceptance in time, including the geometry of the setup.
Once estimated, as demonstrated in this publication, the SF(iso-flux) is expected to improve the so far published results by DB and RENO, providing an extra flux error reduction by up to $\sim$50\% and $\sim$40\%, respectively on $\theta_{13}$.
Due to the simplified conditions assumed for our calculations, our SF's are expected to be slightly optimistic relative to those to be obtained by dedicated analyses by DB and RENO eventually.
%Thus, the final SF is to be estimated by each collaboration in dedicated analyses and detailed publications.
In the case DC-II, the SF(iso-flux) term is expected to yield a dramatic $\sim$90\% error reduction since the iso-flux condition is almost met.
This makes DC-II the only $\theta_{13}$ experiment likely to benefit from a negligible flux error, compared to other systematics.
DC has officially prospected a conservative 0.1\% as flux error for DC-II~\cite{Abe:2013sxa}, well within the analysis here presented.
%that all 1-reactor period (\textgreater30\% of total fraction) will integrate to null flux error.
Therefore, DC-II final $\theta_{13}$ sensitivity is expected to be dominated its challenging background systematic,thus in maximal complementary to DB error budget.
%, due to its shallowest overburden, hence its measurement of 
%The RENO improvement via SF(iso-flux) is expected to have little impact, since it is lowest in magnitude and the background systematic is still large.
Since the flux error is dominant for DB, the hereby presented error reduction by $\sim$50\% translates into a significant improvement of the world $\theta_{13}$ precision by means of DB alone, but also via the envisaged combination by all reactor experiments.
%, benefiting much from the complementary error budgets of DB and DC, with RENO being in between.

\begin{figure}[htb!]
\centering
\includegraphics[scale=0.45]{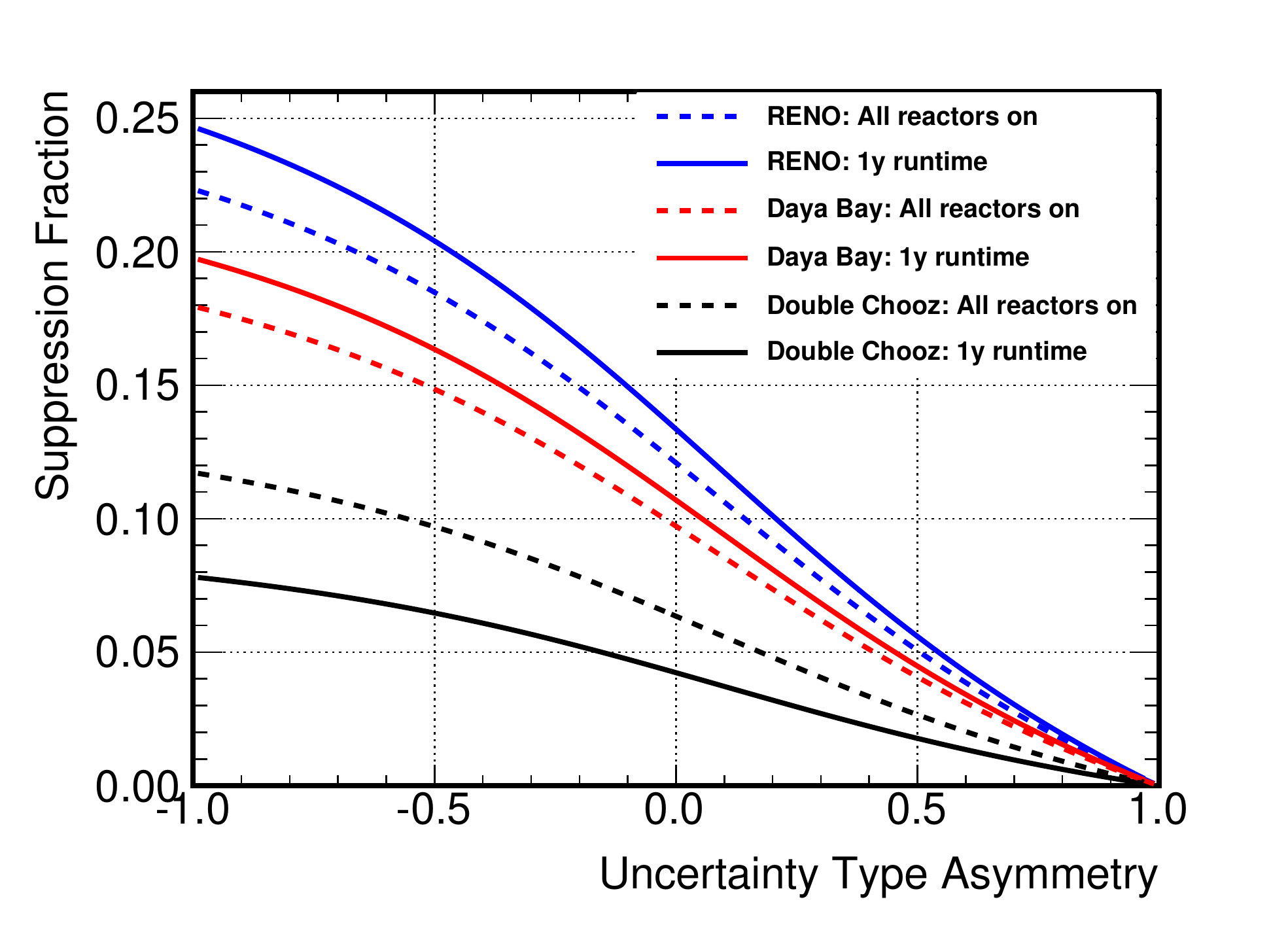}
\caption{
	The variation of SF to the reactor uncertainty type asymmetry, defined as $(\delta^c - \delta^u)/(\delta^c + \delta^u)$, is shown for DC-II, RENO and DB. 
	All experiments have assumed, so far, their errors to be reactor uncorrelated; i.e. $(\delta^c - \delta^u)/(\delta^c + \delta^u)=-1.0$.
	DC-II benefits the best SF ($\sim$0.1) due to its almost iso-flux site.
	RENO and DB benefit mainly from the large number of reactor error suppression, but they also have some partial iso-flux matching hence benefiting from an extra up to $\sim$50 error suppression so far neglected in $\theta_{13}$ publications.
	Both full-reactor power (dashed lines) and a simplified refuelling scenario (solid lines) are shown, while the latter is expected to be a more accurate description of the reality.
	In the refuelling scenario, DC-II benefits from the null reactor flux error whenever only one reactor is running, while RENO and DB have the expected opposite trend.
	}
\label{Fig-ALL_1D}
\end{figure}

The SF(correlation) term has the potential to provide further flux systematic error suppression, regardless of the reactor site geometry, in the (unexpected) limit of full correlation across reactors, SF$\to$0 since the SF(correlation) term cancels, as illustrated in Fig.~\ref{Fig-ALL_1D}.
%(i.e. uncertainty type asymmetry, defined as $(\delta^c-\delta^u)/(\delta^c+\delta^u)$, tends to 1.0), SF(correlation) can lead to full error cancellation.
Assessing the correlation among reactor errors is a difficult subject, having today no specific common prescription or consensus.
Thus the only existing consensus is to adopt the most conservative scenario, implying two distinct cases:

\begin{description}

\item[Single-Detector Case:] 
	maximal SF(total) is obtained if {\it total correlation of reactor errors} is assumed, as shown in Fig.~\ref{Fig-DC_0near_2D}.
	This is because both SF(correlation) and SF($N_R$) terms provide no error suppression; i.e. SF=1 each.
	This the scenario is assumed for all DC-I publications and all single detector experiments, unless otherwise proved, thus affecting all past and and most future experiments.

\item[Multi-Detector Case:]
	maximal SF(total) is obtained if {\it total uncorrelation of reactor errors} is assumed, as shown in Fig.~\ref{Fig-DC_1near_2D}-Right.
	This implies SF(correlation)=1 and SF($N_R$)=$1/\sqrt{N_R}$, hence benefiting from maximal SF($N_R$) reduction.
	This is the scenario expected to be assumed by DB, DC-II and RENO, until otherwise proved.

\end{description}

Of course, reactor errors are unlikely to be neither totally correlated or totally uncorrelated.
The fact that only those extreme cases are considered in the literate is a mere demonstration of the lack of knowledge for a better handling.
Beyond the current debate among experts on the subject, the promising exploitation of SF(correlation) will require strong reactor-type dependences to be accounted and justified carefully by each experiment in dedicated publications.
This means that each experiment will have to analyse their respective reactors, thus providing insight evidence of the error correlation behaviour.
The thermal power contribution, typically indicated by $P_{th}$, depends mainly on the uncertainty analysis of the internal reactor instrumentation data used for power estimation, as provided by the reactor running company.
Instead, the fission fraction evolution, indicated by $\alpha_{f}$, depends mainly on the simulation uncertainties analysis, including the assumptions and input parameters (fuel configuration, etc) used for the modelling and time evolution.
Both the instrumentation and simulations are very specific to each reactor type and, therefore, to the each experiment.
Therefore, dedicated analyses are needed by each experiment to justify the delicate exploitation of SF(correlation), in the same way that this publication aims to illustrates the realisation of the unprecedented SF(iso-flux) exploitation.
% on the precision $\theta_{13}$, for example.

As a consequence of the unsettled complications behind the assessment of SF(correlation) for each experiment, our description here remains generic in its description, as shown in Fig.~\ref{Fig-ALL_1D}, nonetheless, we illustrate and quantify the rationale behind for error suppression and its promising exploitation potential.
%, upon the necessary studies published.
Our approach is also consistent with the fact that none of the $\theta_{13}$ experiments have so far provided detailed publications on the non-trivial reactor systematics analyses, including the quantitative justification of all assumptions used so far.
In the context of the $\theta_{13}$ experiments, a few references exist DC~\cite{Onillon,DRAGON} and DB\cite{DBDRAGON} (nothing yet available for RENO) illustrating a fraction of their reactor flux studies, but not dealing with the inter-reactor error correlation needed for the exploitation of SF(correction).
DC is, however, finalising a dedicated publication~\cite{DCreactor} on their reactor flux systematics quoted so far.
% used to improve the $\theta_{13}$ precision.
Thus, reactor flux error critical subject, despite its major impact to the final precision on $\theta_{13}$, remains unfortunately somewhat obscure in today's literature.
%So, we strongly encourage DC, DB and RENO to exhaustively cover the reactor flux systematics analyses in dedicated publications, as the world $\theta_{13}$ precision depends strongly on this.
As time goes, the statistical error of reactor experiments is no longer dominant, the treatment of the systematics should be clearly laid well in advance to maximise the stability and reliability of the $\theta_{13}$ measurement, whose impact has critical implication transcending reactor neutrino results, affecting, for example, current searches for neutrino CP-violation.

\section{Conclusions}
%%%%%%%%%%%%%%%%%%%%%%%%

This publication provides reactor neutrino experiments with a coherent treatment for reactor flux systematic uncertainty for multi-detector experiments in multi-reactor contributions.
We started with careful treatment of the single detector in a single reactor site scenario, being the most relevant case for most reactor experiments beyond $\theta_{13}$.
We have demonstrated that the challenging reactor flux systematic do not trivially cancel by the adoption of that multi-detector experiments.
However, we have identified several means for error suppression in the context of Double Chooz, Daya Bay and RENO experiments, using simplified scenarios to maximise relative comparison.
We computed an integral error {\it suppression fraction} (SF), which can be broken down into three components, defined as SF(total)$=$SF(iso-flux)$\times$SF($N_R$)$\times$SF(correlation), where 
SF($N_R$) suppresses the uncorrelated error of identical reactors,
SF(iso-flux) suppresses the total error if the site geometry meets fully, or partially, the iso-flux condition 
%(acceptance relation between detectors and reactors)
and SF(correlation) suppresses the error if the reactor errors are fully correlated.
SF(iso-flux) and SF(correlation) could lead to total cancellation of the flux error.
However, only SF($N_R$) has been exploited to improve the $\theta_{13}$ precision, although total cancellation is impossible.
%Our studies embody an important step towards the improvement of the precision of reactor measurements based on two main observations.

This publication deals in detail on the calculation for SF(iso-flux), thus paving the ground for its exploitation, yielding two important observations.
First, DC, once in its near+far configuration, is the only experiment expected to benefit from a negligible reactor flux error, thanks to the $\sim$90\% iso-flux error suppression.
Second, Daya Bay and RENO could also benefit from their partial iso-flux, thus yielding up to $\sim$50\% flux error suppression.
Thus, this publication embodies a major improvement in the global precision of $\theta_{13}$ by improving the precision of all experiments measuring it, including current results.
Finally, we have highlighted the potential for a mechanism, currently neglected, for error suppression relying on further reactor error correlation insight, characterised by the SF(correlation) term, further improving the precision of all multi-detector experiments.

\section*{Acknowledgments}
We would like to thank H.~de~Kerret (IN2P3/CNRS-APC, France) as well as C.~Buck (MPIK, Germany), L.~Camilleri (Columbia University, USA), R.~Carr (Columbia University, USA), L.~Giot (Subatech, France) and M.~Ishitsuka (Tokyo Institute of Technology, Japan) for suggestions and comments on the manuscript. 
This work was supported by the IEF Marie Curie programme (P.~Novella).

%%%%%%%%%%%%%%%%%%%%%%%%

%%%%%%%%%%%%%%%%%%%%%%%%%%%

%%%%%%%%%%%%%%%%%%%%%%%%
\appendix
\section*{Appendix: Error Propagation for Partially Correlated Uncertainties}
\setcounter{section}{1}
%%%%%%%%%%%%%%%%%%%%%%%%

In order to deliver the correlation coefficient for partially correlated uncertainties, we generalise the approach presented in~\cite{Drosg}. 
Keeping the same notations, let us consider a measurement $F$ which depends on two variables $x$ and $y$ having the absolute uncertainties: $\Delta x$ and $\Delta y$. 
We split one of them, for example $\Delta x$, into its two components $\Delta x^u$ and $\Delta x^c$ representing respectively the totally uncorrelated and totally correlated both relative to $\Delta y$. 
Since the general error propagation formula is symmetric and the correlation is mutual, one can split whether $\Delta x$ or $\Delta y$. 
The fraction between the correlated and uncorrelated components is given characterised by a constant $a$ such that $\Delta x^u = a \Delta x^c
$. 
Since $(\Delta x)^2 = (\Delta x^u)^2 + (\Delta x^c)^2$, the uncorrelated and correlated components are expressed as

\begin{equation}     
\Delta x^u = \frac{a \Delta x}{\sqrt{a^2 + 1}} \hspace{0.5cm} \mathrm{and} \hspace{0.5cm} \Delta x^c = \frac{\Delta x}{\sqrt{a^2 + 1}}.
\end{equation}

\noindent
In order to estimate the total uncertainty

\begin{equation}     
(\Delta F)^2 = (\Delta F^u)^2 + (\Delta F^c)^2
\label{eq_apx_Fph1}
\end{equation}

\noindent we shall compute each of its components.
The uncorrelated component is

\begin{eqnarray}     
(\Delta F^u)^2 &=& \left( \frac{\partial F}{\partial x} \right)^2 (\Delta x^u)^2 \nonumber \\
               &=& \frac{a^2}{a^2 + 1}  \left( \frac{\partial F}{\partial x} \right)^2 (\Delta x)^2 
\label{eq_apx_Fu}
\end{eqnarray}

\noindent while the correlated component is
 
\begin{eqnarray}        
(\Delta F^c)^2 &=& \left( \frac{\partial F}{\partial x} \right)^2 (\Delta x^c)^2 + \left( \frac{\partial F}{\partial y} \right)^2 (\Delta y)^2 + \nonumber \\ 
&& + 2 \left( \frac{\partial F}{\partial x} \right) \Delta x^c \left( \frac{\partial F}{\partial y} \right) \Delta y  \nonumber \\
&=& \frac{1}{a^2 + 1} \left( \frac{\partial F}{\partial x} \right)^2 (\Delta x)^2 + \left( \frac{\partial F}{\partial y} \right)^2 (\Delta y)^2 + \nonumber \\
&& + \frac{2}{\sqrt{a^2 + 1}} \left( \frac{\partial F}{\partial x} \right) \Delta x \left( \frac{\partial F}{\partial y} \right) \Delta y
\label{eq_apx_Fc}
\end{eqnarray}
\\
%\vspace{5cm}

\noindent
Replacing Eq.~\ref{eq_apx_Fu} and Eq.~\ref{eq_apx_Fc} in Eq.~\ref{eq_apx_Fph1}, we obtain

\begin{eqnarray}        
(\Delta F)^2 &=& \left( \frac{\partial F}{\partial x} \right)^2 (\Delta x)^2 + \left( \frac{\partial F}{\partial y} \right)^2 (\Delta y)^2 + \nonumber \\
&& + \frac{2}{\sqrt{a^2 + 1}} \left( \frac{\partial F}{\partial x} \right) \Delta x \left( \frac{\partial F}{\partial y} \right) \Delta y
\label{eq_apx_F}
\end{eqnarray}

\noindent
Comparing Eq.~\ref{eq_apx_F} to the general law of error propagation

\begin{eqnarray}        
(\Delta F)^2 &=& \left( \frac{\partial F}{\partial x} \right)^2 (\Delta x)^2 + \left( \frac{\partial F}{\partial y} \right)^2 (\Delta y)^2 + \nonumber \\
&& + 2 k \left( \frac{\partial F}{\partial x} \right) \Delta x \left( \frac{\partial F}{\partial y} \right) \Delta y
\end{eqnarray}

\noindent we define the {\it correlation error coefficient} as

\begin{equation}        
k = \frac{1}{\sqrt{a^2 + 1}}
\label{eq_kdea}
\end{equation}

\noindent
Replacing now the $a$ parameter, we can rewrite the previous equation as

\begin{equation}        
k = \frac{1 + \frac{\Delta x^c - \Delta x^u}{\Delta x^c + \Delta x^u} }{\sqrt{2 \left[1 + \left( \frac{\Delta x^c - \Delta x^u}{\Delta x^c + \Delta x^u} \right)^2 \right]}}
\label{eq_kdebinter}
\end{equation}

\noindent
Eq.~\ref{eq_kdebinter} remains the same for the relative uncertainty

\begin{equation}        
k = \frac{1 + \frac{\delta^c - \delta^u}{\delta^c + \delta^u} }{\sqrt{2 \left[1 + \left( \frac{\delta^c - \delta^u}{\delta^c + \delta^u} \right)^2 \right]}}
\label{eq_kdeb}
\end{equation}

\noindent where $\delta^{u,c}$ stands for $\Delta x^{u,c}/x$, respectively.


\begin{thebibliography}{99}
%%%%%%%%%%%%%%%%%%%%%%%%%%%

\bibitem{Abe:2013sxa}
	Y.~Abe~{\it et al.}~[Double Chooz Collaboration], Improved Measurements of the Neutrino Mixing Angle $\theta_{13}$ with the Double Chooz Detector,
	\textit{JHEP} {\bf 10} (2014) 086.

%\bibitem{DB1} 
%	F.~P.~An~{\it et al.}~[Daya Bay Collaboration], Observation of electron-antineutrino disappearance at Daya Bay,
%	\textit{Phys.\ Rev.\ Lett.} {\bf 108} (2012) 171803

\bibitem{DB2}
	F.~P.~An~{\it et al.}~[Daya Bay Collaboration], Improved Measurement of Electron Antineutrino Disappearance at Daya Bay,
	\textit{Chin.\  Phys.\ C} {\bf 37} (2013) 011001

\bibitem{DB3}
	F.~P.~An~{\it et al.}~[Daya Bay Collaboration], Spectral Measurement of Electron Antineutrino Oscillation Amplitude and Frequency at Daya Bay,
	\textit{Phys.\ Rev.\ Lett.\ } {\bf 112} (2014) 061801

\bibitem{Ahn:2012nd}
	J.~K.~Ahn~{\it et al.}~[RENO Collaboration], Observation of Reactor Electron Antineutrino Disappearance in the RENO Experiment,
	\textit{Phys.\ Rev.\ Lett.\ } {\bf 108} (2012) 191802

\bibitem{MIextra} 
	P.~Adamson~{\it et al.}~[MINOS Collaboration], Measurement of Neutrino and Antineutrino Oscillations Using Beam and Atmospheric Data in MINOS
	\textit{Phys.\ Rev.\ Lett.} {\bf 110} (2013) 251801

\bibitem{T2Kextra} 
	K.~Abe~{\it et al.}~[T2K Collaboration], Observation of Electron Neutrino Appearance in a Muon Neutrino Beam,
 	\textit{Phys.\ Rev.\ Lett.} {\bf 112} (2014) 061802

\bibitem{GFA1extra} 
	M.C.~Gonzalez-Garcia, M.~Maltoni, Th.~Schwetz~[Global Fit Analysis], Updated fit to three neutrino mixing: status of leptonic CP violation,
 	\textit{JHEP} {\bf 1411} (2014) 052

\bibitem{GFA2extra} 
	D.V.~Forero, M.~Tortola, J.W.F.~Valle~[Global Fit Analysis], Neutrino oscillations refitted,
 	\textit{Phys.\ Rev.\ D} {\bf 90} (2014) 093006

\bibitem{GFA3extra} 
	F.~Capozzi, G.L.~Fogli, E.~Lisi, A.~Marrone, D.~Montanino, A.~Palazzo [Global Fit Analysis], Status of three-neutrino oscillation parameters, circa 2013,
 	\textit{Phys.\ Rev.\ D} {\bf 89} (2014) 093018

%\bibitem{DCLorenz}
%	Y.~Abe {\it et al.} [Double Chooz Collaboration], First test of Lorentz violation with a reactor-based antineutrino experiment,
%	\textit{Phys.\ Rev.\ D } {\bf 86} (2012) 112009

\bibitem{prospect}
	Y.-F. Li~{\it et al.}, Unambiguous determination of the neutrino mass hierarchy using reactor neutrinos,
	\textit{Phys.\ Rev.\ D } {\bf 88} (2013) 013008

\bibitem{spectra1} 
	Th.~Mueller~{\it et al.}, Improved predictions of reactor antineutrino spectra,
	\textit{Phys.\ Rev.\ C} {\bf 83} (2011) 054615

\bibitem{spectra2} 
	P.~Huber, Determination of antineutrino spectra from nuclear reactors,
	\textit{Phys.\ Rev.\ C} {\bf 84} (2011) 024617

\bibitem{ILL1}
	K. Schreckenbach~{\it et al.}, Determination of the Antineutrino Spectrum from $^{235}U$ Thermal Neutron Fission Products up to 9.5 MeV
	\textit{Phys.\ Lett.} {\bf 160B} (1985) 325

\bibitem{ILL2}  
	A. A. Hahn~{\it et al.}, Antineutrino Spectra from $^{241}Pu$ and $^{239}Pu$ Thermal Neutron Fission Products
	\textit{Phys.\ Lett.} {\bf 218B} (1989) 365

\bibitem{Djurcic}
	Z. Djurcic {\it et al.}, Uncertainties in the anti-neutrino production at nuclear reactors, 
	\textit{J.Phys.G: Nucl. Part. Phys.} {\bf 36} (2009) 045002

\bibitem{Onillon}   
%{\bf METTRE TITRE EN ANGLAIS : A. Onillon, Pr\'{e}diction des taux de fission reacteur et estimation des incertitudes associ\'{e}es dans le cadre de l’exp\'{e}rience Double Chooz, \textit{PhD Thesis, University of Nantes} (2014). }
	A. Onillon, Prediction of the Reactor Fission Rates and the Estimation of the Associated Uncertainties in the Frame of the Double Chooz Experiment, 	PhD Thesis, University of Mines Nantes (2014) [\cor{}\url{https://tel.archives-ouvertes.fr/tel-01082405}]

\bibitem{DBspent}
	An Feng-Peng {\it et al.}, Systematic impact of spent nuclear fuel on $\theta_{13}$ sensitivity at reactor neutrino experiment 
	\textit{Chinese Phys. C} {\bf 33} (2009) 711

\bibitem{Drosg}
	M.~Drosg, Dealing with Uncertainties - A Guide to Error Analysis, Second Enlarged Edition
	\textit{ISBN 978-3-642-01383-6, Springer-Verlag Berlin Heidelberg} (2009), 152-154. 

\bibitem{DRAGON}
	C.L.~Jones~{\it et al.}, Reactor Simulation for Antineutrino Experiments using DRAGON and MURE.
	%\textit{Chinese Phys. C} {\bf 33} (2009) 711
	{\it Phys. Rev. D} {\bf 86} 012001 (2012).

\bibitem{DBDRAGON}
	X.B.~Ma~{\it et al.}, Uncertainties analysis of fission fraction for reactor antineutrino experiments using DRAGON.
	arXiv:1405.6807
	%\textit{Chinese Phys. C} {\bf 33} (2009) 711

\bibitem{DCreactor}
	Y.~Abe~{\it et al.}~[Double Chooz Collaboration], Reactor Flux Systematic Error for the Double Chooz Experiment.
	{\it In Preparation}.

%\bibitem{DayaBay:Proposal}
%  Daya Bay Collaboration, A Precision Measurement of the Neutrino Mixing Angle $\theta_{13}$ using Reactor Antineutrinos at Daya Bay,
%  \textit{ arXiv:hep-ex/0701029 }


\end{thebibliography}
\end{document}